\documentclass[12pt]{article} 
\pdfoutput=1
\usepackage[utf8]{inputenc}
\usepackage[T1]{fontenc}
\usepackage{lmodern, amsmath, amssymb, slashed, graphicx, comment,adjustbox,bm}
\usepackage[dvipsnames]{xcolor}
\usepackage[numbers, elide, compress]{natbib}
\usepackage[colorlinks=true, citecolor=blue, urlcolor=blue, linkcolor=blue, breaklinks=true, pdfpagelabels=false]{hyperref}

\usepackage[normalem]{ulem}

\makeatletter\g@addto@macro\bfseries{\boldmath}\makeatother

\def\figureautorefname~#1\null{Fig.\,#1\null}

\def\equationautorefname~#1\null{Eq.\,(#1)\null}

\numberwithin{equation}{section}

\newcommand{\infb}{\,{\rm fb}^{-1}}

\newcommand{\eehz}{e^+e^- \to hZ}

\newcommand{\eeww}{e^+e^- \to W^+W^-}

\newcommand{\eett}{e^+e^- \to t\bar{t}}

\newcommand{\bpm}{\begin{pmatrix}}
\newcommand{\epm}{\end{pmatrix}}

\usepackage{pifont}
\begin{document}

\begin{flushright}
\end{flushright}

\vspace*{1.5cm}

\begin{center}

{\Large\bf
From Optimal Observables to Machine Learning: an Effective-Field-Theory Analysis of $\eeww$ at Future Lepton Colliders
\par}
\vspace{9mm}

{\bf Shengdu~Chai$\,^{a}$},~~~{\bf Jiayin~Gu$\,^{a,b}$},~~~{\bf Lingfeng Li$\,^{c}$}\\ [4mm]
{\small\it
$^a$ Department of Physics and Center for Field Theory and Particle Physics, \\ Fudan University, Shanghai 200438, China \\[2mm]
$^b$ Key Laboratory of Nuclear Physics and Ion-beam Application (MOE), \\ Fudan University, Shanghai 200433, China  \\[2mm]
$^c$ Department of Physics, Brown University, Providence, RI, 02912, USA
\par}
\vspace{.5cm}
\centerline{\tt \small sdchai19@fudan.edu.cn,  jiayin\_gu@fudan.edu.cn, lingfeng\_li@brown.edu}

\end{center}

\vspace{1cm}

\begin{abstract}
We apply machine-learning techniques to the effective-field-theory analysis of the $\eeww$ processes at future lepton colliders, and demonstrate their advantages in comparison with conventional methods, such as optimal observables.  Compared to traditional algorithms, we show that simulation-based inference methods are more robust to detector effects and backgrounds, and could in principle produce unbiased results with sufficient Monte Carlo simulation samples that accurately describe experiments. This is crucial for the analyses at future lepton colliders given the outstanding precision of the $\eeww$ measurement ($\sim 10^{-4}$ in terms of anomalous triple gauge couplings or even better) that can be reached.  Our framework can be generalized to other effective-field-theory analyses, such as the one of $\eett$ or similar processes at muon colliders.
%
%
\end{abstract}

\newpage
{\small 
\tableofcontents}

\setcounter{footnote}{0}

\section{Introduction}

Future lepton colliders have great physics potentials, and have been extensively studied by the particle physics communities~\cite{Narain:2022qud}. 
These include several versions of circular and linear $e^+e^-$ colliders~\cite{CEPCPhysicsStudyGroup:2022uwl, Bernardi:2022hny, ILCInternationalDevelopmentTeam:2022izu, CLICdp:2018cto, Bai:2021rdg}, as well as muon colliders which are technologically more challenging but could reach a higher center-of-mass energy~\cite{Aime:2022flm, Accettura:2023ked}.  A major goal of these colliders is to precisely measure the Higgs boson couplings~\cite{deBlas:2019rxi, deBlas:2022aow, Forslund:2022xjq}.   Processes such as $e^+e^- \to f\bar{f}$ (both on and off the $Z$-pole), $\eeww$ and $\eett$ will also be measured.  While they are important on their own, they also provide crucial inputs for the global analyses in the framework of the Standard Model Effective Field Theory (SMEFT), and have nontrivial interplay with both the Higgs measurement and each other~\cite{Falkowski:2015jaa, Durieux:2017rsg, Barklow:2017suo, Durieux:2018tev, Durieux:2018ggn, Ellis:2018gqa, DeBlas:2019qco, Durieux:2019rbz, Ellis:2020unq, Liu:2022vgo, deBlas:2022ofj, Ethier:2021bye, Brivio:2022hrb, Bartocci:2023nvp, Allwicher:2023shc,Wen:2023xxc}. 

For many processes, including $\eeww$ and $\eett$, it is essential to extract the information from differential distributions, which are sensitive to the new physics contributions. Different methods vary in terms of complexity and effectiveness ({\it i.e.} how much information is extracted). One of the simplest approaches is to bin the distribution and combine the likelihood (chi-square) from the rate measurements in all bins.\footnote{Indeed, the binned-distribution method has been used in many SMEFT analyses of $\eeww$~\cite{ Falkowski:2015jaa, Durieux:2017rsg, Ellis:2018gqa, Marchesini:2011aka, Bian:2015zha, Grojean:2018dqj, Subba:2022czw, Subba:2023rpm}.} However, it becomes impractical if the phase space of the process has a large dimension. For example, assuming both $W$s are on shell, each $\eeww$ event can be characterized by five angles, one for the production and two from each W decay. A 5-dimensional binned distribution can have too many bins for a practical analysis, while projecting the distribution to a lower phase-space dimension necessarily throws away part of the information.   The other extreme is to construct the likelihood from all the measurement events directly, given that one could calculate the differential cross section (in terms of the model parameters), which is essentially the probability density function up to normalization. This method is often denoted as the matrix-element method, as one needs to first calculate the matrix element. It could, in principle, extract the maximum amount of information from data and provide the best statistical reach, but it can be time and computational-power-consuming.  

There also exists a method that is both simple and effective, but with the requirement that the observables depend on (new physics) parameters only at the linear level~\cite{Diehl:1993br}. In this case, one only needs to measure a set of so-called ``optimal observables'', from which the full likelihood (as in the matrix-element method) can be constructed, and the method is thus statistically optimal. More specifically, assuming the differential cross section is of the form
\begin{equation}
 \frac{d \sigma}{d \Omega} = S_0 +\underset{i}{\sum} S_{1,i} \,g_i \,,  \label{eq:dsoo} 
\end{equation} 
where $S_0$ and $S_1$ are functions of the phase space $\Omega$, and $g_i$ are a set of physics parameters, it was shown in Ref.~\cite{Diehl:1993br} that the inverse covariance matrix of the $g_i$ corresponding to the optimal reach are given by 
\begin{equation}
c^{-1}_{ij} = \int d\Omega \frac{S_{1,i}S_{1,j} }{S_0} \cdot \mathcal{L} \,,
\end{equation}
where $\mathcal{L}$ is the integrated luminosity. The $c^{-1}_{ij}$ can be obtained from the optimal observables, defined as $\mathcal{O}_i = \frac{S_{1,i}}{S_0}$, with the relation $c^{-1}_{ij} = n V_{ij}$ where $V_{ij}$ is the covariance matrix of the optimal observables and $n$ is the number of events.  The central values of $g_i$ are given by
\begin{equation}
E[g_i] = \sum_j c^{-1}_{ij} ( E[\mathcal{O}_j]  - E_0[\mathcal{O}_j]) \,,
\end{equation} 
where $E$ is the expectation and $E_0$ is the SM expection with all $g_i$ set to zero. In the SMEFT framework, assuming the new physics scale $\Lambda$ is sufficiently larger than the electroweak (EW) scale and the energy scale of the experiment, the leading new physics contribution to observables is given by the interference term between the SM amplitude and the one of dimension-6 operators.  Keeping only this leading contribution, and discarding the sub-leading contributions such as the dimension-6 squared terms and the interference between SM and dimension-8 operators, observables can be written in terms of \autoref{eq:dsoo} where $S_0$ is the SM differential cross section and $g_i$ are the dimension-6 Wilson coefficients.  Importantly, this linear approximation gives a very good parameterization of almost all the Higgs, EW and top measurements at future lepton colliders, as a result of the outstanding precision that can be achieved.\footnote{On the contrary, it may not be the case for the measurements at LHC, especially  at the high energy tails.  See {\it e.g.} Refs.~\cite{Contino:2016jqw, Alte:2017pme} for more discussions.}

Both the matrix-element method and the optimal-observable one rely on accurate knowledge of the differential cross section for the measured process, which can not be achieved in practice. Detector effects, backgrounds, as well as initial and final state radiations, all need to be taken into account. While some of these effects can be calculated (to some extent) and convoluted into the differential cross section, doing so  would easily generate very complicated expressions for the differential cross section, making the calculation impractical. Generally speaking, there is a distinction between the ``truth-level'' differential cross section (for which one could calculate an analytic expression) and the measured one. For the latter, one could at best have a sample from Monte Carlo simulation, assuming it accurately simulates all the effects mentioned above.\footnote{In practice, one also needs to take account of the theory and parametric uncertainties from missing higher-order loop contributions or measurements of input parameters. We do not address this issue in this paper.} One may argue that, with the clean environment at future lepton colliders, systematic effects are subdominant, and the truth-level differential cross sections would be very good approximations of the measured ones. However, for exactly the same reason,  lepton colliders can reach a high measurement precision, demanding the precision of theoretical predictions to be at least at a similar level. As we will show explicitly later, an unacceptable bias in the reach of the Wilson coefficients could be introduced if (small) systematic effects are not taken into account. It is therefore desirable to extract a ``detector level'' differential cross section, or equivalent statistical quantities such as the likelihood ratios, directly from the Monte Carlo samples.

This goal, while difficult for conventional methods, has become increasingly achievable in recent years with the rapid development of machine learning techniques in high energy physics~\cite{Brehmer:2018kdj, Brehmer:2018eca, Brehmer:2018hga, Brehmer:2019xox, DAgnolo:2019vbw, Chen:2020mev, Chen:2023ind, Brehmer:2019gmn, Butter:2021rvz, Chatterjee:2021nms, GomezAmbrosio:2022mpm, Arganda:2022qzy, Chatterjee:2022oco, Grojean:2022mef, Alasfar:2022vqw, Letizia:2022xbe, Li:2020vav, Yang:2021kyy, Yang:2022fhw, Dong:2023nir, Metodiev:2017vrx, Nachman:2021yvi, Gambhir:2022dut, Gambhir:2022gua, Feickert:2021ajf,  dAgnolo:2021aun, Arganda:2022zbs, Arganda:2023qni} (see also Refs.~\cite{Guest:2018yhq, Carleo:2019ptp,Cranmer:2019eaq,Karagiorgi:2021ngt, ParticleDataGroup:2022pth} for reviews of the subject). With powerful computing tools and enough training samples, a neural network can be trained as a good estimator of the detector-level likelihood ratios, which can be used on real data to obtain the optimal reaches on the model parameters (Wilson coefficients), at least in the limit of large MC statistics and perfect training. Furthermore, the Monte Carlo tools also allow one to exploit the truth-level information that could greatly improve the training efficiency and results~\cite{Brehmer:2018kdj, Brehmer:2018eca, Brehmer:2018hga}.

In this paper, we apply machine learning methods to a well-studied process at lepton colliders, $\eeww$.  Recent SMEFT analysis using optimal observables have shown that the measurements of this process at future colliders can reach a precision of $\sim10^{-4}$ or even better in terms of anomalous triple gauge couplings (aTGCs), under somewhat idealistic assumptions~\cite{DeBlas:2019qco, deBlas:2022ofj}.  We consider the tree-level process with contributions from CP-even dimension-6 operators and focus on the semi-leptonic channel.  On top of the truth-level process, we apply detector simulations and include $e^+e^-\to ZZ$ as a possible background.  We compare the performances of the optimal observables and machine learning methods, and in particular, a machine-learning version of the optimal observables (denoted as SALLY in \cite{Brehmer:2018kdj, Brehmer:2018eca}) and its variations.  We found that, while at the parton level, the optimal observables give the optimal reach and the machine-learning methods have no advantage, with detector effects and backgrounds, the optimal observables easily generate an unacceptable bias, while a much smaller bias (subject to limited MC training samples and imperfect training) can be obtained with appropriate machine-learning methods.  This demonstrates, as proof of principle, that machine-learning methods are much more robust under detector effects and backgrounds, which can be crucial for future collider analyses.   


%


The rest of this paper is organized as follows:   In \autoref{sec:eeww} we provide a brief overview of the theoretical and phenomenological aspects of the $\eeww$ process.  In \autoref{sec:ml}, we lay out the framework of our machine learning analyses.  The details of the sample preparation and training are described in \autoref{sec:simulation}.  Our results are presented in \autoref{sec:result}, which are based on a circular collider running at 240\,GeV.  Finally, we conclude in \autoref{sec:con}. In \autoref{app:chis}, we provide the numerical $\chi^2$ for the full EFT parameterization (excluding modifications of $m_W$) for several scenarios considered in \autoref{sec:result}.  In \autoref{app:nn8}, we present the results of the individual neutral network models which provides a measure of the systematic error due to imperfect training and illustrates the importance of the averaging step described in \autoref{sec:simulation}.

\section{The $\eeww$ process}
\label{sec:eeww}

%
\begin{figure}[t]
\centering
\includegraphics[height=2.5cm]{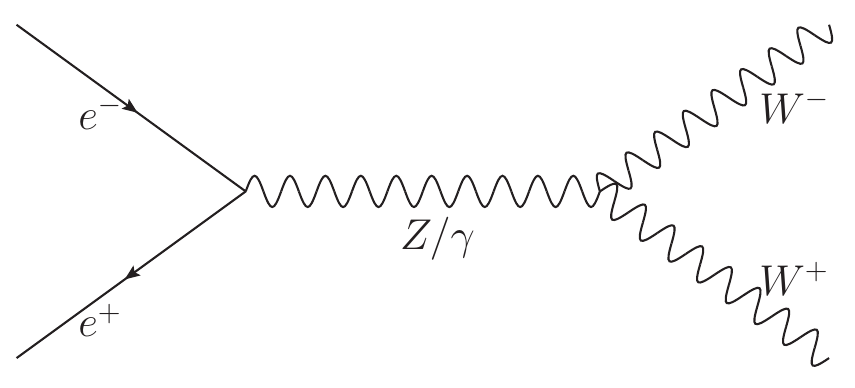} \hspace{1cm}
\includegraphics[height=2.5cm]{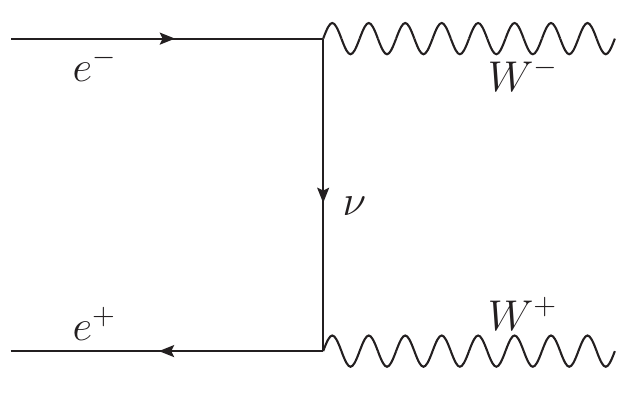} 
\caption{Tree-level Feynman diagrams of the $\eeww$ process.}
\label{fig:feynww}
\end{figure}
Measuring $\eeww$ process is an important part of the precision EW and Higgs program at future Lepton colliders.  Conventionally, the new physics contributions are parameterized in terms of the aTGCs, assuming that they enter only the 3-gauge-boson vertex in the left diagram of \autoref{fig:feynww}.  In a more general framework, all contributions from dimension-6 operators in SMEFT should be considered~\cite{Zhang:2016zsp}.  Focusing on CP-even tree-level effects, contributions from dimension-6 operators to $\eeww$ can be written in terms of effective couplings, including the 3 independent aTGCs, $\delta g_{1Z}$, $\delta\kappa_\gamma$ and $\lambda_Z$, 
\begin{align}  
{\mathcal L}_{\rm TGC} = ~& i e  (W_{\mu\nu}^+ W^{-\mu} - W_{\mu\nu}^- W^{+\mu}  ) A^\nu + i e (1+ \delta \kappa_\gamma) A^{\mu\nu} W_\mu^+ W_\nu^-  \nonumber \\
+& \, i g  c_w \left[ (1+\delta g_{1Z}) (W_{\mu\nu}^+ W^{-\mu} - W_{\mu\nu}^- W^{+\mu})Z^\nu + (1+\delta g_{1Z}- \frac{s^2_w}{c^2_w} \delta \kappa_\gamma)\, Z^{\mu\nu} W_\mu^+ W_\nu^- \right]  \nonumber\\
+& \, \frac{ig \lambda_Z}{m_W^2} \left( s_w W_{\mu}^{+\nu} W_{\nu}^{-\rho} A_{\rho}^{\mu} +c_w W_{\mu}^{+\nu} W_{\nu}^{-\rho} Z_{\rho}^{\mu} \right)\,,   \label{eq:atgc}
\end{align}
and modifications to the electron gauge couplings $\delta g^\ell_W$, $\delta g^e_{Z,L}$ and $\delta g^e_{Z,R}$ (or the muon ones for muon colliders),
\begin{align}  
{\mathcal L}_{Vff} =&  - \frac{g}{\sqrt{2}}  (1+\delta g^\ell_W ) \left[ W^+_\mu \bar{\nu}_L \gamma^\mu e_L + {\rm h.c.} \right] \nonumber \\
& - \frac{g}{c_w} Z_\mu \left[ \bar{e}_L \gamma^\mu (-\frac{1}{2}+s^2_w+\delta g^e_{Z,L} ) e_L + \bar{e}_R \gamma^\mu (s^2_W+\delta g^e_{Z,R} ) e_R  \right] + \ldots \,,
\end{align}
where $s_w \equiv \sin\theta_W$, $c_w \equiv \cos\theta_W$ and $\theta_W$ is the weak mixing angle.  Dimension-6 operators could also modify the $W$-boson mass, $m_W$.  Here we simply set $m_W$ to be SM-like, given that it will be measured to a precision of $\mathcal{O}(10^{-5})$ (at the MeV level) at future lepton colliders.  Note that a large set of operators could modify the $W$ branching ratio.  While they need to be included in a more general global-fit framework, here we ignore their effects for simplicity.  Practically, this can be done by including only the differential information and discarding the likelihood (chi-squared) of the total rate measurement since modifications of the $W$ branching ratio only affect the latter.  With these assumptions, the new physics effects in $\eeww$ can be fully captured by the following six parameters:
\begin{equation}
\delta g_{1Z}  \,, ~~~~~  \delta \kappa_\gamma   \,, ~~~~~ \lambda_Z   \,, ~~~~~  \delta g^\ell_W   \,, ~~~~~  \delta g^e_{Z,L}   \,, ~~~~~  \delta g^e_{Z,R} \,.
\label{eq:para6}
\end{equation}
Our machine-learning analysis in the next section is based on this 6-parameter framework.  Furthermore, in \autoref{sec:result} we mainly focus on the 3 aTGCs, $\delta g_{1Z}$, $\delta \kappa_\gamma$ and $\lambda_Z$, assuming other parameters are already sufficiently constrained by Z-pole and W (width and branching ratio) measurements.  
It is straightforward to extend the framework and include more parameters, including coefficients of CP-odd operators.  However, the focus of this paper is on the machine learning methods rather than providing the most general global-fitting results. 

\begin{figure}
    \centering
    \includegraphics[width = 0.8\linewidth]{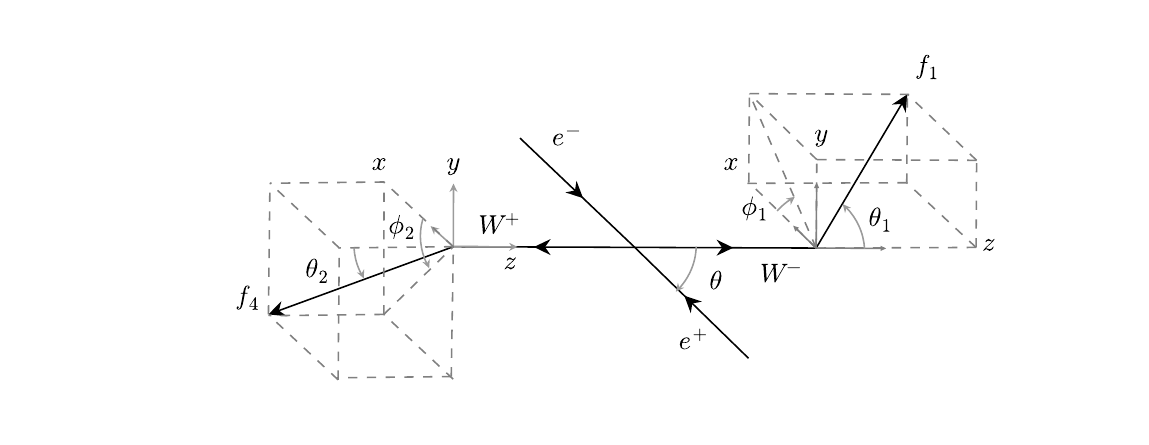}
    \caption{Assuming the $W$ bosons are on shell, each $WW$ event can be characterized by 5 angles, including the production polar angle $\theta$ and 2 angles from each $W$ decay.  Note that $\theta_1$ and $\theta_2$ are defined in the rest frames of $W^-$ and $W^+$, respectively.}
    \label{fig:5angles}
\end{figure}

We adopt the narrow width approximation (NWA) and assume both $W$ bosons are on shell. As such, each $WW$ event can be characterized by five angles, as shown in \autoref{fig:5angles}. $\theta$ is the production polar angle between the incoming $e^-$ and the outgoing $W^-$.\footnote{As the standard notation in statistics, the symbol $\theta$ is also used to denote model parameters later in \autoref{sec:ml}.  The meaning of $\theta$ should be clear from the context.}  $\theta_1$ is the angle between $f_1$ and the $W^-$ (with $W^- \to f_1 \bar{f}_2$, where $f_1=\ell^-, d,c$) in the rest frame of $W^-$, $\phi_1$ is the angle between the decay plane and the production plane formed by $e^-,e^+,W^-,W^+$; For $W^+$, a similar pair of angles $\theta_2$, $\phi_2$ can also be defined with $f_3=\ell^+,\bar{d},\Bar{c}$. The differential cross section, $\frac{d\sigma}{d\Omega} \equiv \frac{d\sigma}{d\cos\theta d\cos\theta_1 d\phi_1 d\cos\theta_2 d\phi_2}$, is universal for all $W$ decay channels at the parton level with massless quarks and leptons.  However, for hadronic decays, the strong QCD shower creates multiple hadrons and forms jets, making the two quarks highly indistinguishable in practise~\footnote{The quantum number of the jet-initiating quark still leaves imprints after hadronization. Several techniques, such as jet flavor tagging and jet charge (see {\it e.g.} Refs.~\cite{Subba:2022czw, Subba:2023rpm, Kamenik:2023ytu, Fraser:2018ieu} have already been studied for the LHC and future lepton colliders.). These evidences are, in general, insufficient to confirm the nature of the jet-initiating quark. Currently, such detailed jet information is not included in our analysis for simplicity and conservativeness, but it can be incorporated within our framework easily.}. For the rest of this work, the corresponding differential cross sections are ``folded'' by taking the average of two different quark configurations. 
We also found the impact of NWA to be minimal from our numerical result. This is due to the difference between NWA and full matrix element results largely cancel after taking joint likelihood ratios. 

\section{The framework of machine learning analysis}
\label{sec:ml}
\subsection{Simulation-based inference by machine learning}
\label{ssec:framework}


In this section, we will introduce the simulation-based (or equivalently likelihood-free) inference~\cite{Cranmer:2019eaq} framework of our machine learning analysis. We first list here the essential statistical quantities and the notations we use for them, which mainly follow the ones in Refs.~\cite{Brehmer:2018kdj, Brehmer:2018eca,Brehmer:2018hga}. 
For simplicity, we will often use the unbolded symbols $\theta$, $x$ and $z$ instead of $\bm{\theta}$, $\mathbf{x}$ and $\mathbf{z}$, and they still generally denote a column of quantities rather than a single one.  
\begin{itemize}

    \item $\bm{\theta}$ is the set of parameters of interest, which we wish to determine from experiments. In SMEFT, they are simply the Wilson coefficients of the higher dimensional operators. More explicitly, in our study, we have
\begin{equation}
\bm{\theta} = \{\, \delta g_{1Z}  \,,  \delta \kappa_\gamma   \,, \lambda_Z   \,, \delta g^\ell_W   \,,   \delta g^e_{Z,L}  \,,  \delta g^e_{Z,R} \, \}  \,.
\end{equation}
For most of the results presented in \autoref{sec:result}, we will focus on the 3-aTGC parameterization, $\bm{\theta} = \{ \, \delta g_{1Z} \,,  \delta \kappa_\gamma   \,, \lambda_Z \, \}$.  

    \item $\mathbf{x}$ is the set of detector-level observables that we can measure. In principle, one could have a very large set of observables at multiple levels, including the low-level observables such as detector readout, {\it etc.}  Here we focus on the particle-level observables,  assuming some basic particle reconstructions ({\it e.g.} jet clustering) has already been done. These include the IDs (if any) and 4-momenta of visible particles, as well as observables constructed from them, such as the reconstructed 5 angles. For a given event $e$, the values of $\mathbf{x}$ are denoted as $\mathbf{x}_e$.

    \item $\mathbf{z}$ is the set of truth-level observables, including the ID and 4-momenta of all particles and the observables constructed from them. They cannot be measured in experiments, and are called ``latent variables'' in Refs.~\cite{Brehmer:2018kdj, Brehmer:2018eca,Brehmer:2018hga}. However, in MC simulations, we have access to them (which are just the parton level events) and also know the mapping between $\mathbf{x}_e$ and $\mathbf{z}_e$ for any given event.  

    \item $p(z|\theta)$ is the truth-level probability density function for a given parameter $\theta$.  It is the differential cross section normalized by the total cross section, $\frac{1}{\sigma} \frac{d \sigma}{d \Omega}$. This can be computed analytically, at least in principle. Given a set of $N$ events, we can construct a total 
    likelihood from $p(z|\theta)$, given by $\overset{N}{\underset{e=1}{\Pi}} p(z_e|\theta)$, with $z_e$ the values of observables of the event $e$.  In practice, it is more convenient to use the log likelihood, which turns the product into a sum. 

    \item $p(x|\theta)$ is the detector-level probably density function.  
    Formally, it can be convoluted from the truth-level probably density function as
\begin{equation}
p(x|\theta)=\int dz p(x|z) p(z|\theta),  \label{eq:factor}
\end{equation} 
where $p(x|z)$ contains the information on the parton showering, detector simulations, {\it etc.}  Due to the stochastic nature of the generative process from $\mathbf{z}\to \mathbf{x}$, it is generally not possible to directly calculate $p(x|\theta)$. 
For $N$ events one could construct a total likelihood, $\overset{N}{\underset{e=1}{\Pi}} p(x_e|\theta)$.

    \item $p(x,z|\theta) = p(x|z)p(z|\theta)$ is the joint probability density of $x$ and $z$.

    \item $r(x|\theta_0, \theta_1)=\frac{p(x|\theta_0)}{p(x|\theta_1)}$ is the likelihood ratio between two hypothesis, $\theta_0$ and $\theta_1$.  According to the Neyman-Pearson lemma, the likelihood ratio provides the optimal statistical test for a binary hypothesis problem.  By custom, $\theta_1$ usually denotes the SM, with all Wilson coefficients of higher dimensional operators set to zero.  For a given set of events, we can directly construct the total likelihood ratio between $\theta_0$ and $\theta_1$ as $\overset{N}{\underset{e=1}{\Pi}} r(x_e|\theta_0, \theta_1)$, if $p(x|\theta)$ is known. 
    However, since $p(x|\theta)$ is generally not calculable, neither is $r(x|\theta_0, \theta_1)$ or $\overset{N}{\underset{e=1}{\Pi}} r(x_e|\theta_0, \theta_1)$.

    \item $r(x,z|\theta_0, \theta_1)=\frac{p(x,z|\theta_0)}{p(x,z|\theta_1)}=\frac{p(x|z)p(z|\theta_0)}{p(x|z)p(z|\theta_1)} = \frac{p(z|\theta_0)}{p(z|\theta_1)}$ is the joint likelihood ratio of $x$ and $z$.  Note that we assume $p(x|z)$ is universal for all events and thus cancels in the expression, making this quantity calculable with relatively low computational cost. Following the discussion in~\cite{Brehmer:2018eca}, the knowledge on the joint-likelihood $r(x,z|\theta_0, \theta_1)$ is able to determine $r(x|\theta_0, \theta_1)$ with sufficient data, which will be explained in detail below.

    \item For a given test sample with $N$ events, one could also include the information of the total rate in the likelihood ratio.  Assuming the total rates follow Poisson distributions, the total likelihood ratio can be written as 
 %
\begin{equation} 
R(X|\theta_0, \theta_1) \equiv \overset{N}{\underset{e=1}{\Pi}} r(x_e|\theta_0, \theta_1) 
 = \frac{\overset{N}{\underset{e=1}{\Pi}} p(x_e|\theta_0) \cdot \frac{n_0^N e^{-n_0}}{N!} }{\overset{N}{\underset{e=1}{\Pi}} p(x_e|\theta_1) \cdot \frac{n_1^N e^{-n_1}}{N!} }  
 = \frac{\overset{N}{\underset{e=1}{\Pi}} d\sigma(x_e|\theta_0) \cdot e^{-n_0}}{\overset{N}{\underset{e=1}{\Pi}} d\sigma(x_e|\theta_1) \cdot  e^{-n_1} } \,,
\end{equation} 
where $X$ denotes collectively all the detector-level observables in the full data set, $n_0(\theta_0)$ and $n_1(\theta_1)$ are the expected number of events for hypothesis $\theta_0$ and $\theta_1$. $d\sigma\equiv \frac{d\sigma}{d\Omega}$ is the differential cross section.  (Note that $n_0 = \sigma(\theta_0) \cdot \mathcal{I}$ and $n_1 = \sigma(\theta_1) \cdot \mathcal{I}$, where the integrated luminosity $\mathcal{I}$ cancels in the ratio).  It is then convenient to define a modified likelihood ratio~\cite{Chen:2020mev},
\begin{equation} 
\tilde{r}(x|\theta_0, \theta_1)\equiv\frac{d\sigma(x|\theta_0)}{d\sigma(x|\theta_1)} \,,
\end{equation} 
and similarly the modified joint likelihood ratio, $\tilde{r}(x,z|\theta_0, \theta_1)\equiv\frac{d\sigma(z|\theta_0)}{d\sigma(z|\theta_1)}$.  Note that $d\sigma(z)$ and $d\sigma(x)$ correspond to the parton-level and detector-level differential cross sections, respectively. The modified (joint) likelihood ratio $\tilde{r}$ differs from the unmodified $r$ only by a factor of $\sigma(\theta_0)/\sigma(\theta_1)$. The total log likelihood ratio for $N$ events can then be written as~\footnote{For a Bayesian analysis, the posterior needed is $R(\theta_0,\theta_1|X)\equiv p(\theta_0|X)/p(\theta_1|X)$ instead. However, similar to many other precision tests of particle physics, the prior $p(\theta)$ we adopt is uniform/flat, making $R(\theta_0,\theta_1|X)$ identical to $R(X|\theta_0,\theta_1)$ after constant factors cancel out.}
\begin{equation} 
\log R(X|\theta_0, \theta_1) = \overset{N}{\underset{e=1}{\sum}} \log \tilde{r}(x_e|\theta_0,\theta_1) +n_1 - n_0.
\end{equation} 

\end{itemize}
In reality, we do not have access to $p(x|\theta)$, but only MC simulated events sampled with $p(x|\theta)$. Our goal is to obtain an estimator of the (modified) likelihood ratio, $\hat{r}(x|\theta_0, \theta_1)$ or $\hat{\tilde{r}}(x|\theta_0, \theta_1)$,\footnote{Note that the estimators of $n_1$ and $n_0$ can be easily obtained from simulation.} from a set of (or multiple sets of) MC samples which we call the training samples.    Ideally, this estimator should closely resemble the true likelihood ratio. This is done by constructing a so-called ``loss function(al)'' $L$, which is a functional of the estimator $\hat{r}(x|\theta_0, \theta_1)$ or some variations of it. A required feature of the loss function is that it is minimized by the true likelihood ratio in the large statistics limit. Many ways of constructing the loss function exist, some of which rely only on MC samples. However, in MC simulation, the parton level events sampled by $d\sigma(z|\theta)$ are also available and can be exploited to increase the training efficiency, as pointed out in Refs.~\cite{Brehmer:2018kdj, Brehmer:2018eca, Brehmer:2018hga}.  This requires the knowledge of $d\sigma(z|\theta)$ and the mapping between $x_e$ and $z_e$, both of which are accessible from MC simulations.

To obtain a good estimator for $r(x|\theta_0,\theta_1)$, we are interested in the loss functional of the form~\cite{Brehmer:2018eca}
\begin{equation}
\label{eq:proof1}
L[\hat{r}(x|\theta_0,\theta_1)]=\iint dx dz ~p(x,z|\theta_1)\bigg[r(x,z|\theta_0,\theta_1)-\hat{r}(x|\theta_0,\theta_1) \bigg]^2~.
\end{equation}
The loss functional reaches minimum if the estimator $\hat{r}(x|\theta_0,\theta_1)$ satisfies:
\begin{equation}
\frac{\delta L}{\delta \hat{r}(x|\theta_0,\theta_1)} = 2\int dz ~p(x,z|\theta_1) \bigg[r(x,z|\theta_0,\theta_1)-\hat{r}(x|\theta_0,\theta_1) \bigg]=0~,    
\end{equation}
which indicates
\begin{align}
\label{eq:proof3}
&\int d z ~p(x,z|\theta_1) r(x,z|\theta_0,\theta_1) = \int d z ~p(x,z|\theta_1) \frac{p(x,z|\theta_0)}{p(x,z|\theta_1)} =p (x|\theta_0) \\\nonumber
&= r(x|\theta_0,\theta_1)p(x|\theta_1) =  r(x|\theta_0,\theta_1)\int dz ~p(x,z|\theta_1) = \hat{r}(x|\theta_0,\theta_1) \int dz ~p(x,z|\theta_1)~.
\end{align}
Therefore, a proper estimator $\hat{r}(x|\theta_0,\theta_1)= r(x|\theta_0,\theta_1)$ minimizes the loss functional $L$ after the latent space $z$ is ``integrated out''.  The argument above is still valid when replacing $r$'s with $\hat{r}$'s as they only differ by a factor independent of $x$ and $z$.

In practice, the integration of $\iint dx dz ~p(x,z|\theta_1)$ from ~\autoref{eq:proof1} is represented by the MC generation/sampling procedure with parameter $\theta_1$. The loss functional $L$ is recognized as the loss function evaluated event-by-event correspondingly. In the context of machine learning, one is able to construct a general ansatz for $\hat{r}(x|\theta_0, \theta_1)$ and obtain a good estimator by numerically minimizing the loss function of a set of training samples with respect to the neural network parameters, which is often referred to as the training process. 

In the most general case, each training only gives us the likelihood ratio of two points ($\theta_0$ and $\theta_1$) in the model parameter space. However, if additional assumptions are made, for instance, $p(x|\theta)$ is linear or quadratic in $\theta$, then only a finite set of training is needed to obtain an estimator $\hat{r}(x|\theta, \theta_1)$ for the entire parameter space of $\theta$. In SMEFT analyses, a quadratic dependence of the Wilson coefficients is generally a good assumption and has been implemented in {\it e.g.} Ref.~\cite{Chen:2020mev, GomezAmbrosio:2022mpm}. Furthermore, with the large statistics and high precision of the $\eeww$ measurement at future lepton colliders, it is a very good approximation to keep only the leading linear contributions of the dim-6 Wilson coefficients, in which case the likelihood ratio and the joint likelihood ratio could be simply written as  
\begin{align}
\label{eq:linearexpansion}
     \tilde{r}( x|\theta_{0} ,\theta_{1}) = \frac{d\sigma(x|\theta_0)}{d\sigma(x|\theta_1)} =1+\sum _{i=1}^{n} \alpha_{i}(x)( \theta _{0,i} -\theta _{1,i}) \,, \\
    \tilde{r}( x,z|\theta_{0} ,\theta_{1}) = \frac{d\sigma(z|\theta_0)}{d\sigma(z|\theta_1)} =1+\sum _{i=1}^{n} \alpha_{i}(z)( \theta _{0,i} -\theta _{1,i}) \,, \label{eq:rmodlinear}
\end{align}
where $n$ is the number of model parameters (dim-6 Wilson coefficients) and $\alpha _{i}(z)$ are just the parton-level optimal observables. Once we obtain an estimator $\hat{\alpha}_{i}(x)$ for each $\alpha_{i}(x)$, we could construct an estimator $\hat{\tilde{r}}( x|\theta  ,\theta_{1})$ for the entire parameter space $\bm{\theta}$.  We could thus directly construct a loss function in terms of the $\hat{\alpha}_{i}(x)$. 
By plugging in \autoref{eq:rmodlinear} and replacing the integration $\iint dx dz ~p(x,z|\theta_1)$ with the sum over the events sampled with $p(x,z|\theta_1)$, the loss function in \autoref{eq:proof1} can be written as
\begin{equation}
\iint dx dz ~p(x,z|\theta_1)  \bigg[r(x,z|\theta_0,\theta_1)-\hat{r}(x|\theta_0,\theta_1) \bigg]^2  \to \sum_e  \bigg[ \sum _{i=1}^{n} \left( \alpha_{i}(z_e) -  \hat{\alpha}_{i}(x_e) \right) ( \theta _{0,i} -\theta _{1,i}) \bigg]^2 \,.
\end{equation}
One could then obtain a loss functional for each $\hat{\alpha}_i$ of the form $\sum_e |\hat{\alpha}_i(x_e)-\alpha_i(z_e)|^2$ by choosing a suitable set of $\theta_{0,i}$ ({\it i.e.} by switching on one Wilson coefficient at a time).  Finally, the $n$ loss functionals could simply be summed together into a more compact form, which is minimized when each one is minimized.  We therefore have 
\begin{equation}
L(\hat{\alpha}_i) =  \sum_{i, e}|\hat{\alpha}_i(x_e)-\alpha_i(z_e)|^2, \label{eq:losally}
\end{equation}
%
%
%
where, by default,  $z_e$ and $x_e$ are always sampled according to $p(x,z|\theta_1)$.
\autoref{eq:losally} is minimized for $\hat{\alpha}_i(x_e) = \alpha_i(x_e)$ in the large statistics limit. The dependence on $(\mathbf{\theta}_0-\mathbf{\theta}_1)$ is dropped out as it is independent of $x$ and $z$. The method using the loss function in \autoref{eq:losally} is denoted as the Sally (Score Approximates Likelihood Locally) method in Ref.~\cite{Brehmer:2018eca}, where the $\alpha_i$s are the ``scores.'' As we noted above, it is essentially a machine-learning version of the optimal observable analysis, and is the main method that we use. We have compared it with several other methods and found Sally to give the best result in our analysis, which is expected since the linear approximation (for the dependence on Wilson coefficients) works very well in our case.  


Finally, with an estimator $\hat{\tilde{r}}(x|\theta, \theta_1)$,\footnote{Note that we have replaced $\theta_0$ by $\theta$ here to denote that $\hat{\tilde{r}}$ is a function of the model parameters $\theta$ instead of a particular set of values of them, $\theta_0$.} for a set of test samples (either from actual experiments or from simulations) one could construct the $\chi^2$ as a function of the model parameters $\theta$, given by 
\begin{equation}
\chi^2(\theta)- \chi^2(\theta_1)\equiv -2 \log \hat{R}(X|\theta,\theta_1) = -2\left( \sum_e\log \hat{\tilde{r}} (x_e|\theta,\theta_1) +n_1-n(\theta) \right) \,, \label{eq:chis}
\end{equation}
which can be used to set limits on the model parameters $\theta$.  
$\Delta\chi^2(\theta) = \chi^2(\theta) - \chi_{\rm min}^2(\theta)$ can be obtained by first finding the minimum of \autoref{eq:chis} and then subtract it from \autoref{eq:chis}.



\subsection{Machine Learning with Background events}
\label{subsec:background}
In particle physics, background events are inevitable in most studies for various reasons. In general, background events will be involved in the analysis due to a finite detector resolution that projects events with distinct $z$ representations into the same $x$ vector. In many cases, the region in $z$ where background events dwell may even has no overlap with that of signal events. This case becomes quite generic when the definition of $z$ includes all truth-level information from an MC generator, such as the identities of intermediate states. Moreover, in practice, the simulation of signal and background processes does not have to be within the same framework or even handled by varying MC generators. 

We first recognize the signal and background regions ($z_{\rm sig}$ and $z_{\rm bkg}$) in the $z$ space so that the two regions cover all possible events and do not overlap with each other:
\begin{equation}
\label{eq:region_2}
z_{\rm sig}\cap z_{\rm bkg} = \emptyset~.
\end{equation}
We assume all backgrounds are insensitive to parameter $\theta$ so that $\frac{\partial}{\partial \theta} p(z|\theta)\equiv 0$ when $z \in z_{\rm bkg}$ for simplicity. The likelihood of $p(x|\theta)$ is made up by the signal ($ p(x|\theta)_{\rm sig}$) and background ($ p(x)_{\rm bkg}$) contribution. The $\theta$ dependence of the latter is omitted since $p(z|\theta)$ does not depend on $\theta$. According to the above discussion, the key for simulation-based inference is to predict the likelihood ratio $r(x|\theta_0,\theta_1)$ and its modified counterpart $\tilde{r}(x|\theta_0,\theta_1)$. In the presence of backgrounds, their relationship reads
\begin{align}
    r(x|\theta,\theta_1)
    &= \frac{\sigma(\theta_1)_{\rm sig}+\sigma_{\rm bkg}}{\sigma(\theta)_{\rm sig}+\sigma_{\rm bkg}} \times  \frac{d \sigma(x|\theta)_{\rm sig}+d \sigma(x)_{\rm bkg}}{d \sigma(x|\theta_1)_{\rm sig}+d \sigma(x)_{\rm bkg}}=\frac{\sigma(\theta_1)_{\rm sig}+\sigma_{\rm bkg}}{\sigma(\theta)_{\rm sig}+\sigma_{\rm bkg}}  \tilde{r}(x|\theta,\theta_1)~,
\end{align}
where the subscripts indicate the contribution from $z_{\rm sig}$ or $z_{\rm bkg}$ accordingly. For precision measurements, the expansion in~\autoref{eq:linearexpansion} on $\tilde{r}(x|\theta,\theta_1)$ is modified as
\begin{equation}
\label{eq:backgroundexpansion}
\tilde{r}(x|\theta,\theta_1) = 1+ \frac{d \sigma(x|\theta_1)_{\rm sig} }{d \sigma(x|\theta_1)_{\rm sig}+d \sigma(x)_{\rm bkg}} \sum _{i=1}^{n} \alpha_{i}(x)( \theta _{i} -\theta _{1,i})~.
\end{equation}
The factor $S(x|\theta_1) \equiv\frac{d \sigma(x|\theta_1)_{\rm sig} }{d \sigma(x|\theta_1)_{\rm sig}+d \sigma(x)_{\rm bkg}} \leqslant 1$ modifying $\alpha$ stems from the impact of background events around a certain observable configuration $x$. Its value is not known in general from the MC sampling due to the highly stochastic projection from the latent $z$ space to the observable $x$ space. As a result, the value of $S(x|\theta_1)$ will be evaluated by machine learning methods.

Interestingly, the loss function in~\autoref{eq:losally} can stabilize around the correct output if the training sample is weighted properly. Let's begin with a training sample being a mixture of signal and background events, with their number ratio the same as the ratio of their cross sections in the SM ($\theta_1$). When the sample size is very large, for any coarse-grained observable point $x^\prime$, the total weight of signal and background samples will be proportional to $d\sigma(x^\prime|\theta_1)_{\rm sig}$ and $d\sigma(x^\prime)_{\rm bkg}$, respectively. For the subset of samples around $x^\prime$, the loss function $L(\hat{\alpha})$ takes minimum when
\begin{equation}
    \frac{\partial}{\partial \hat{\alpha}_i}L(\hat{\alpha}_i) \propto 2[ d\sigma(x^\prime|\theta_1)_{\rm sig}(\hat{\alpha}_i(x^\prime) - \alpha_i(x^\prime)) + d\sigma(x^\prime)_{\rm bkg}(\hat{\alpha}_i(x^\prime) - 0)]=0~.
\end{equation} The above equation is satisfied when $\hat{\alpha}_i(x^\prime)=S(x^\prime|\theta_1)\alpha_i(x^\prime) $, which can be directly used to calculate $r(x^\prime|\theta,\theta_1)$. Therefore, the Sally algorithm described in the last section is applicable to cases with non-trivial backgrounds when trained with the correctly weighted samples.

Besides the method above, any other method that reproduces $S(x|\theta_1) \alpha(x)$ as its output is also valid for simulation-based inference. One of the simplest ways is to train a separate classifier model on a mixed sample of signal and background events since $S(x|\theta_1)$ is a typical output of a classifier with a binomial log-likelihood or quadratic loss~\cite{Cranmer:2015bka}. Such a classifier model can be trained separately and work together with a Sally inference model trained only on signal event samples. The product of the two models' output will be the final output for $\hat{R}(X|\theta,\theta_1)$ calculation. 

\section{Simulation and training}
\label{sec:simulation}
\subsection{Preparation of samples}
\label{subsec:sample}

The $\eeww$ signal events can be first classified into three cases depending on how many leptons are generated from the $W$ pair decay. Although the dileptonic channel with both $W$ decaying leptonically is potentially the cleanest one that is largely free from hadronic systematics, it has the smallest branching ratio ($\lesssim 10\%$). In addition, with two missing neutrinos, their momenta can not be reconstructed unambiguously.~\footnote{Assuming both $W$'s are on shell, the neutrino momenta can be obtained by imposing energy-momentum conservation and $W$ on-shell conditions. However, there are often degenerate solutions~\cite{Bian:2015zha}.} Conversely, the fully-hadronic channel with both $W$ decaying to quarks has a large branching ratio but suffers more from hadronic systematics. The charge information of outgoing $W$'s crucial for the analysis is also significantly washed out by QCD effects. In this paper, we focus on the semileptonic channel where only one of the $W$ bosons decays leptonically, which enjoys a large branching ratio, uniquely reconstructed missing momentum, and $W$ charge information kept by the isolated lepton track.

In order to evaluate the performance of different methods when various systematic effects are present, Monte Carlo (MC) simulations of $e^{+}e^{-} \to W^{+}W^{-}\to qq\ell \nu$ events are performed with proper data processing. In particular, we use Madgraph\,5~\cite{Alwall:2011uj} to generate unweighted $e^{+}e^{-} \to W^{+}W^{-}\to qq\ell \nu$ samples at $\sqrt{s}=240\,\mathrm{GeV}$.\footnote{Note that, while this energy is not special for $W^{+}W^{-}$, it is one of the standard benchmarks for future lepton colliders as the cross section of the Higgsstrahlung process, $\eehz$, is maximized around this energy. On the other hand, the sensitivities to the aTGCs are strongly suppressed near the $WW$ threshold, which makes the $240\,$GeV run more suitable for our study than the $WW$ threshold run.}  The initial state radiation (ISR) from the electron beam is also incorporated using the method of Ref.~\cite{Frixione:2021zdp}. The parton shower and hadronization are handled by Pythia\,8~\cite{Sjostrand:2007gs}. Delphes\,3~\cite{deFavereau:2013fsa} is used to incorporate detector effects with the detector profile derived from the ILD~\cite{Chen:2017yel}. A lepton is isolated if the net energy of other particles within its $\Delta R< 0.5$ cone is smaller than 0.12 (electron) or 0.25 (muon) of the lepton energy. All jets are clustered with the anti-$k_t$ algorithm~\cite{Cacciari:2008gp} with $R=0.6$. During the procedure, only SM samples are generated as demonstrated in \autoref{eq:proof1} (where $\theta_1$ denotes the SM). The background samples are simulated in the same way, while in this case, we only consider $ee\to ZZ \to qq\ell\ell$ as the major background for simplicity. Such events will fake the semileptonic $WW$ signal if one of the final state leptons is not identified. The reasons for being unable to identify a lepton may include the low lepton energy below the threshold, large $|\eta|$ away from the detector coverage, or accidental escape from gaps between detector materials. By choosing the $ZZ\to qq\ell\ell$ event as the representative background, we make the assumption that the background is insensitive to the SMEFT dim-6 contributions.  This is not true in reality, but is nevertheless a good approximation as the leading dim-6 contribution to this process effectively only modifies the $Zf\bar{f}$ vertices, which are constrained by Z-pole measurements.\footnote{On a separate note, the contribution of dim-8 operators in the $ZZ$ process has also been extensively studied (see {\it e.g.} Refs.~\cite{Ellis:2019zex,Ellis:2020ljj}).  We do not consider dim-8 operators here.} We claim that the purpose of our simulation is not to derive the potential of future lepton colliders with ML in the most realistic context. Instead, we aim to provide a proof of concept that, with the help of ML, the future lepton colliders can overcome various sources of systematics and reach the target precision, while a full simulation of background events and their EFT dependence will be left to future studies.

For the event selection, we apply only the basic requirements that each event contains exactly two jets and one lepton, and all three particles are required to have $p_T > 5\,$GeV. Finally, all simulated samples are converted to a vector of data as the input for ML methods. In our case, the input will be dimension 21. The first 16 dimensions consist of the four-momenta of outgoing leptons, jets, and neutrinos. Here the neutrino four-momentum is obtained by subtracting the vector sum of the momenta of the two jets and the lepton from the initial state with $\sqrt{s}=240$~GeV. The vector of input also contains the five angles ($\cos\theta, \cos\theta_1, \phi_1, \cos\theta_2, \phi_2$) derived from the momenta above, forming an $N\times(16+5)=21\times N$-dimensional dataset, where $N$ is the number of events. Notice that this is only a fraction of the original information of the whole event, which typically involves tens of outgoing particles. Including more detailed event information is certainly compatible with our framework, resulting in a much larger $x$ space. Since we target validating the machine-learning framework in the context of precision tests of the SM, only basic jet and lepton kinematics are involved to reduce the complexity and cost of network training.

%
%

\begin{figure}
    \centering
    \includegraphics[width=\textwidth]{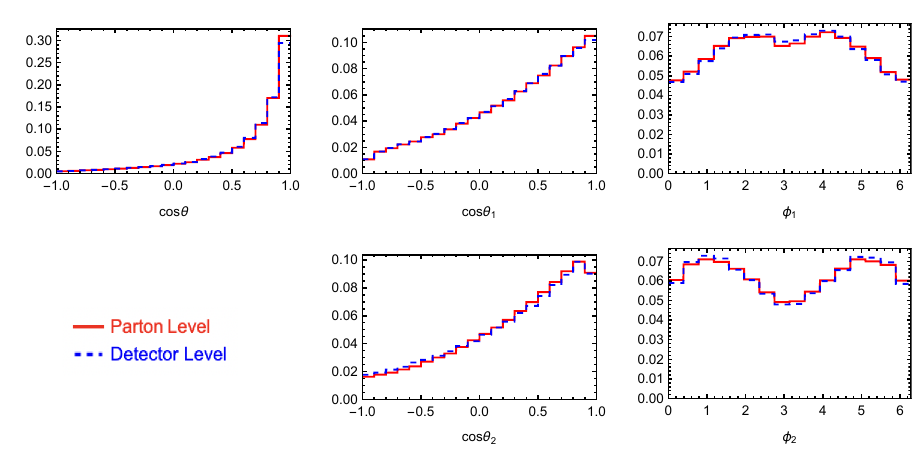}
    \includegraphics[width=\textwidth]{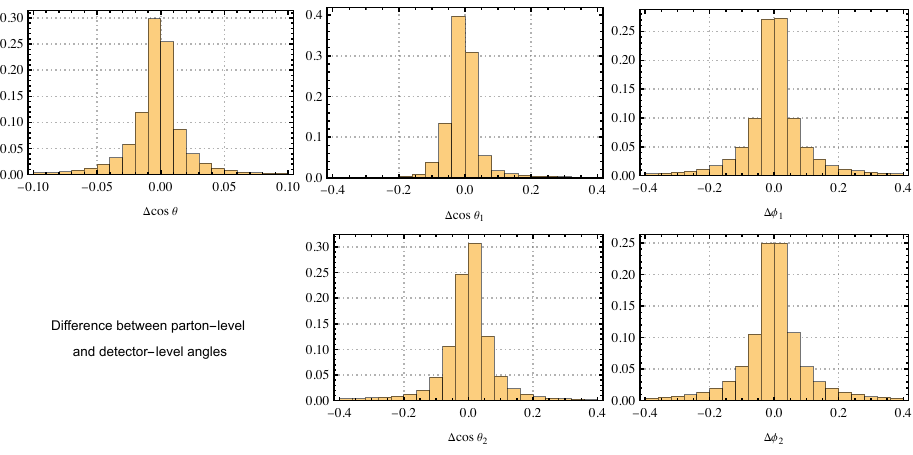}
    \caption{ {\bf Top:} The distributions of the five angles ($\cos\theta, \cos\theta_1, \phi_1, \cos\theta_2, \phi_2$) at the parton level (red) and detector level (blue).  {\bf Bottom:} The difference between parton-level and detector level angles ($\theta_{\rm detector} - \theta_{\rm parton}$) plotted for each angle.}
    \label{fig:angle}
\end{figure}

Given that the key observable in our study is the differential distribution of the five angles ($\cos\theta, \cos\theta_1, \phi_1, \cos\theta_2, \phi_2$), it is important to see how it is modified from the parton level to the detector level.  We show in \autoref{fig:angle} the individual distributions of the five angles at the parton level and the detector level (top panel), as well as the distributions of the difference between the parton level and the detector level angles.\footnote{That is, for each event, we calculate the difference between the parton-level and detector level angles, and then plot the distribution of this difference for all events.  This is done for each of the five angles ($\cos\theta, \cos\theta_1, \phi_1, \cos\theta_2, \phi_2$).  
%
$\cos\theta_2$ and $\phi_2$ always denote the decay angles of the hadronic $W$.  Note that the distribution of $\phi_1$ and $\phi_2$ differ by roughly a phase of $\pi$ due to the convention in their definitions. At the detector (Delphes) level the two jets are indistinguishable.  To obtain the full distribution, we simply match the detector level jets to the parton-level ones (which are distinguishable) with the smaller overall angle differences.  This almost always gives the correct pairing due to the small smearing effects.  
While the smearing effects are small and approximately symmetrical, they can still have nontrivial impacts on the angular distributions.  For instance, the polar angle distributions are slightly flattened.  The effects are barely visible in \autoref{fig:angle}.  However, due the high precision goal of our study, these small effects could have a significant impact on the results if not properly taken account of, as we will show in \autoref{sec:result}.    
}


\subsection{Machine Learning Algorithms and Training Details}

For the need of different strategies, three types of training and testing samples are prepared.
\begin{itemize}

  \item  {\bf Sample-T} ({\bf T}ruth level) is the truth-level sample (before hadronization) with signal only.  

  \item  {\bf Sample-D} ({\bf D}etector level) is the detector-level sample with signal events only.

  \item  {\bf Sample-DB} ({\bf D}etector level with {\bf B}ackground) is the detector-level sample with both the $\eeww$ signal and the $e^+e^- \to ZZ$ background.

\end{itemize}
Each set of training sample has a total number of $2\times 10^6$ events.  For Sample-T and Sample-D, all events are semilepton $WW$ signal.  For Sample-DB, 90\%  are $WW$ signal and 10\% are $ZZ$ background events (which pass the pre-selection). The signal-to-background ratio, 9:1, is slightly larger than the actual one, which we found to be around 15:1 after the preselection cuts.  The ratio 9:1 is chosen for simplicity and being conservative, since our goal is to study the performance of machine-learning methods under background rather than estimating the actual reach.  Each set of validation sample has $5\times 10^5$ $WW$ events, and again for Sample-DB this contain 10\% $ZZ$ events. As mentioned above, we use the Sally method to train the machine-learning models that outputs an estimator of the likelihood ratio.

Three simulation-based inference networks are trained from the three different samples. They are denoted as {\bf Sally-T}, {\bf Sally-D} and {\bf Sally-DB} models, respectively. For all background events, we set $\alpha_i(z_e)=0$ in the loss function in \eqref{eq:losally},
which implies the approximation that background is independent of $\theta$ (and is always SM-like). 
%

%

In {\bf Sally-DB}, the differential cross sections of signal and background are combined with the correct weighting (and assuming background is always SM-like) as described in \autoref{subsec:background}. Conversely, the background can be filtered by a trained classifier network before the inference network. {\bf Sally-DC}  ({\bf D}etector level with {\bf C}lassifer) combines Sally-D and a classifier that filters the background from the signal. The classifier networks is trained separately on the mixed Sample-DB, while the inference network is identical to the Sally-D model with no extra training. 
For comparison, we also consider the {\bf OOC} method, which is optimal observables ({\bf OO}) combined with the {\bf C}lassifier that removes backgrounds.

One of the key motivations of this work is to evaluate various ML strategies for EW precision tests at lepton colliders. A simple fully connected neural network (FCNN) is adopted instead of a more delicate network structure to reduce the training cost. Such a FCNN setup with $\mathcal{O}(10^5)$ parameters is able to handle the 21-dimensional data from semileptonic $\eeww$ events. All inference networks are of 9 layers and 200 nodes per layer for our inference network for Sally algorithms. The input of the neural network is the observable $x$, which has a size of $N\times21$, where $N$ is the number of events in the training set, and the 21 features include the four-momentum of four final-state particles and five angles. The outputs are a set of $\hat{\alpha}_i(x)$, one for each model parameter. For the classifier, we simply change the output of the neural network to be a binary output, which is 1 for signal and 0 for background. In the training process, we use the ADAM optimization~\cite{kingma2017adam} algorithm for stochastic gradient descent and apply early stopping~\cite{caruana2000overfitting} to prevent over-fitting, as well as learning rate decay strategy.




To better evaluate machine learning strategies and in particular, to eliminate the random noise from the training phases, it is beneficial to consider the ensembles of individual models~\cite{dietterich2000ensemble}.  
%
We introduce 
\textbf{Sally-DA} (\textbf{D}etector level \textbf{A}verage), \textbf{Sally-DBA} (\textbf{D}etector level with \textbf{B}ackground \textbf{A}verage) and \textbf{Sally-DCA} (\textbf{D}etector level with \textbf{C}lassifier \textbf{A}verage), which are averaged versions of Sally-D, Sally-DB and Sally-DC, respectively.   
%
%
In each case, we trained 8 neural network models in parallel, which are different from each other in terms of the parameterization and training history. 
In particular, the initial randomization of weights and the random segmentation/batching of the training data are different for the 8 models, while the architecture of all 8 models are the same.
The reconstructed $\hat{\alpha}$ (obtained by minimizing the loss function in \autoref{eq:losally}) of the 8 models are averaged to produce the final result.  Assuming the biases from different models are uncorrelated, the averaging would effectively reduce the model uncertainty.  
As we will show in \autoref{app:nn8}, results of the individual models do exhibits non-negligible systematic error (especially in the central values) which can be effectively reduced by the averaging step.
%


\section{Results}
\label{sec:result}

With the setup in the previous section, we perform machine learning analyses to obtain an estimator for the likelihood ratio from training samples, which is used to construct a $\chi^2$ of the validation sample. For better comparison between different algorithms, when presenting the limits on EFT parameters, the corresponding signal event number for each limit is scaled to $10^4$ events. This is equivalent to a luminosity of $2.2$~fb$^{-1}$. Note that this is only a tiny fraction of all proposed future lepton collider runs of $\mathcal{O}(\text{ab})^{-1}$.

While the $\chi^2$ is obtained for the most general parameterization in \autoref{eq:para6}, we choose to present the results from a global fit of 3 aTGCs, $\{ \delta g_{1Z} \, \delta \kappa_\gamma   \,, \lambda_Z \}$. This is because a global fit with all 6 parameters contains flat directions and is unbounded without the inclusion of other EW measurements, $e.g.$ $Z$-pole observables. Such a 3-aTGC parameterization is well-justified for circular colliders with a tera-Z program, as shown in {\it e.g.} Ref.~\cite{DeBlas:2019qco}. The full $\chi^2$ for each sample type (obtained from the method with the best performance) is provided in \autoref{app:chis}.

\begin{figure}
    \centering
    \includegraphics[height=4.5 cm]{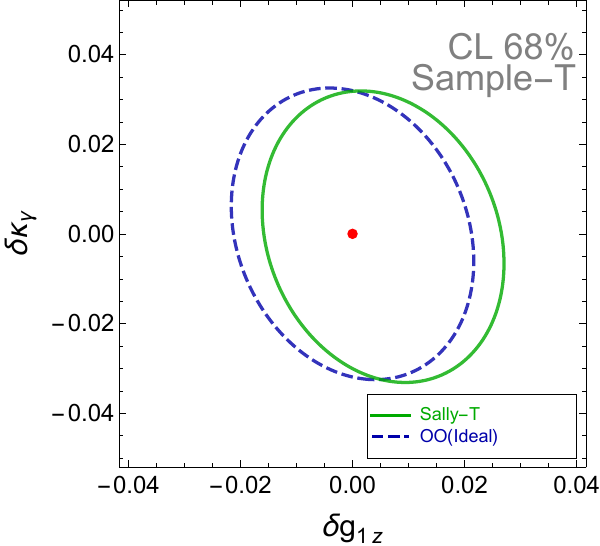}
    \includegraphics[height=4.5 cm]{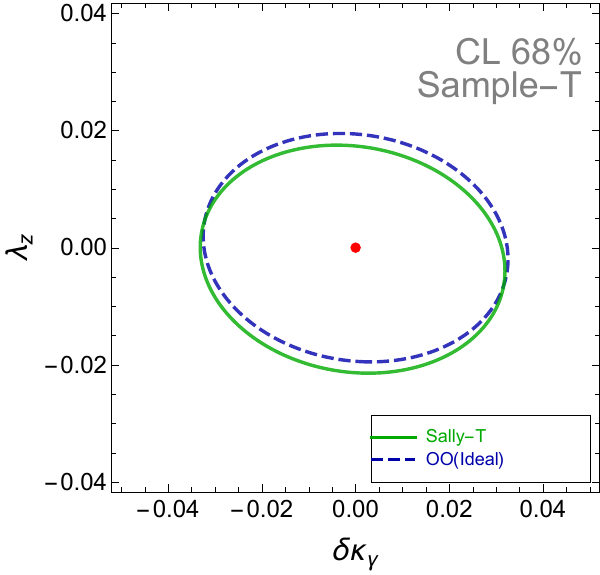}
    \includegraphics[height=4.5 cm]{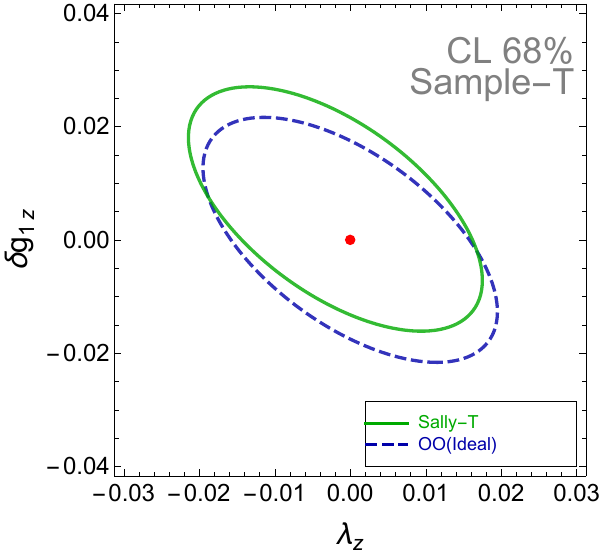}\\
     \includegraphics[height=4.5 cm]{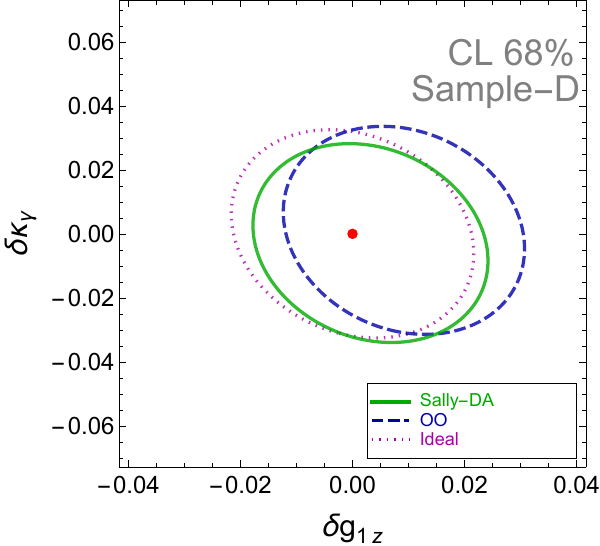}
    \includegraphics[height=4.5 cm]{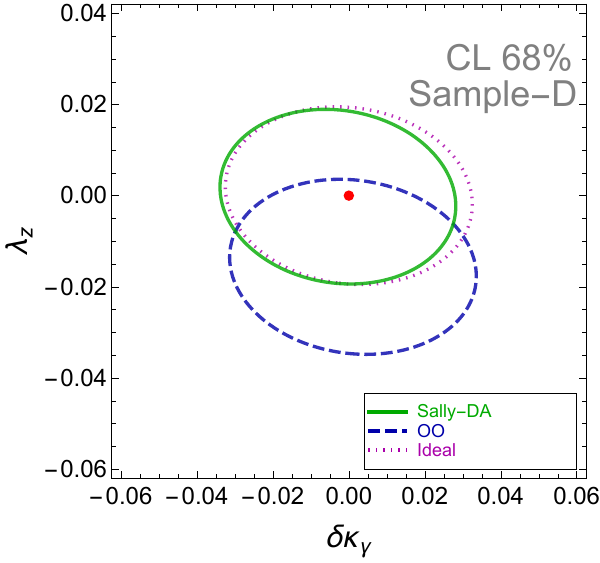}
    \includegraphics[height=4.5 cm]{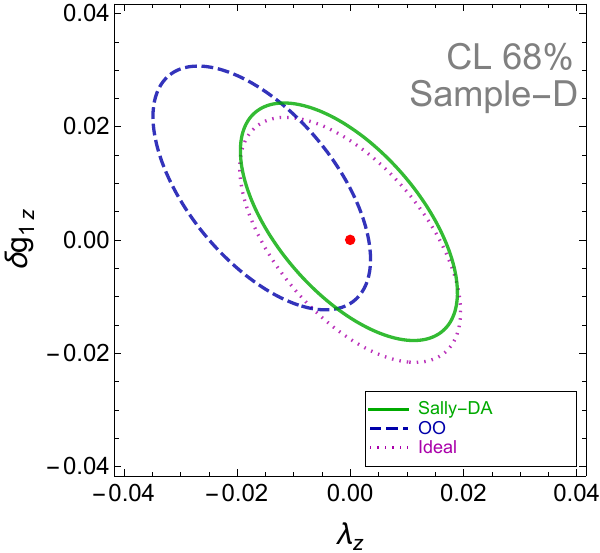}\\
     \includegraphics[height=4.5 cm]{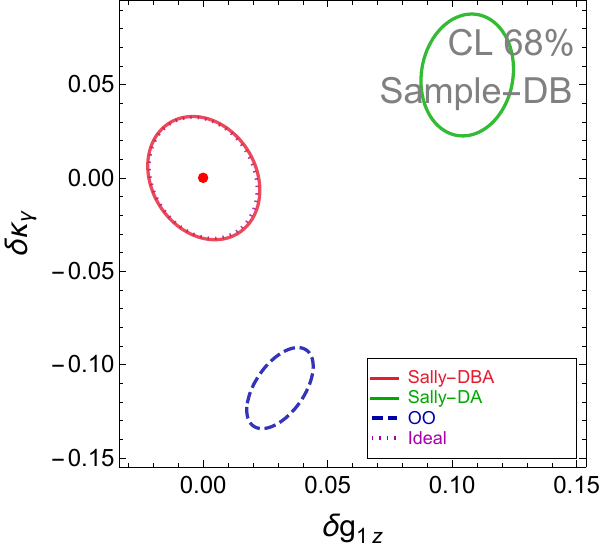}
    \includegraphics[height=4.5 cm]{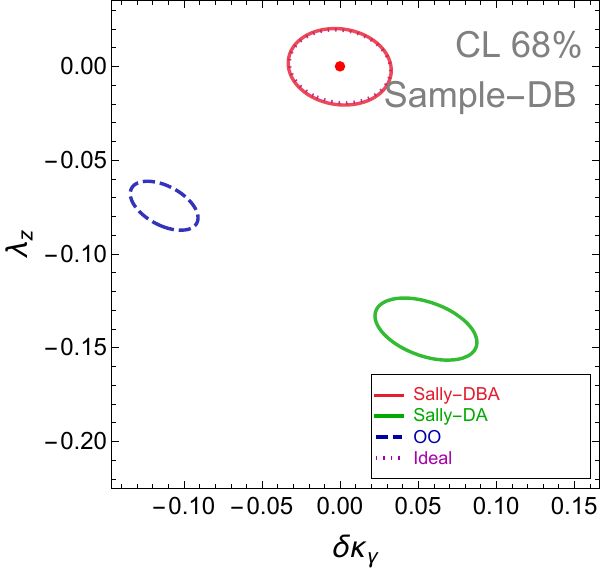}
    \includegraphics[height=4.5 cm]{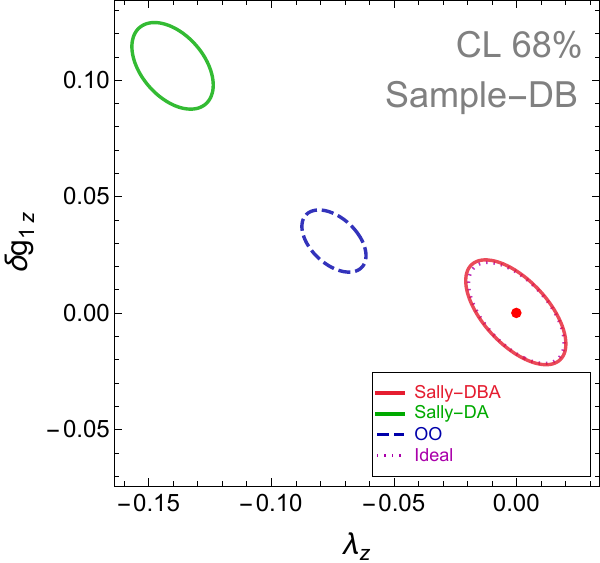}
    \caption{The 68\%CL contours ($\Delta \chi^2=2.28$) in the $(\delta g_{1,Z},\delta \kappa_\gamma)$ (left), $(\delta \kappa_\gamma, \lambda_Z)$ (middle) and $(\lambda_Z, \delta g_{1,Z})$ (right) planes  obtained from the 3-aTGC fit, assuming a signal sample (semi-leptonic $WW$) with $10^4$ events. Each row corresponds to a different sample set, with the results obtained from several methods. \textbf{Top: Sample-T} with Optimal Observables (OO, which coincides with the ideal result in this case) and Sally-T methods. \textbf{Middle: Sample-D} with OO and Sally-D methods. \textbf{Bottom: Sample-DB} with OO, Sally-D and Sally-DBA methods. All contours are calibrated so that the ideal result is always centered at the SM prediction $\delta \kappa_\gamma=\lambda_Z=\delta g_{1Z}=0$. The deviations of each contour's center, therefore, only stem from systematic effects or training-induced quantities instead of MC fluctuations. }
    \label{fig:chi2}
\end{figure}

Our main results are presented in \autoref{fig:chi2}, which shows the limits on the three aTGCs for various scenarios. The results in each row come from the same (3-parameter) global fit, which is projected on the $(\delta g_{1Z}, \delta\kappa_\gamma)$, $(\delta \kappa_\gamma, \lambda_Z)$ and $(\lambda_Z, \delta g_{1Z})$ planes, respectively. Each contour shows the 68\% confidence level (CL) bound for two d.o.f., which corresponds to $\Delta \chi^2=2.28$.  For the first row, the parton-level Sample-T, which corresponds to the most ideal case, is used. Two relevant sets of results are shown, which are the one from applying Optimal Observables (OO, dashed blue), and the one from Sally-T (green). To better illustrate the performance of various algorithms, the minimal $\chi^2$ point ($i.e.$, the center) of the OO result is calibrated to the origin such that $\delta \kappa_\gamma=\lambda_Z=\delta g_{1Z}=0$. This is because OO at the truth level is the ideal inference method. It will not be affected by systematic effects and can thus always recover the truth value.\footnote{Strictly speaking, the statement is only exact when the MC sample size approaches infinity. In practice, it is a good approximation as long as the MC sample size is significantly greater than the experimental yield.} All other contours are evaluated with the same test sample set and calibrated by the same amount as the ideal ones, ensuring that deviations shown are due to systematic effects or training-induced quantities instead of MC fluctuations. In this case, Sally-T has no obvious advantage over OO since it suffers from imperfect training. Nevertheless, we found Sally-T to have a good performance, and its results are close to the ideal ones. 

In the second row of \autoref{fig:chi2}, Sample-D is used, which includes parton shower, hadronization, various detector effects, and ISR radiations. Three sets of contours are presented, corresponding to the ideal limit (the same as the OO result in the first row), the results from Sally-DA, and OO (which is now directly applied on detector-level events). 
For OO, a significant bias in the central value of the fit is observed, which is the result of detector effects that make $p(x|\theta)$ different from $p(z|\theta)$. In particular, the five $\phi$ and $\theta$ angle recovered from detector level information deviates from the truth, leading to inaccurate $\alpha$ predictions. Conversely, Sally-T is replaced by Sally-DA, and its performance only slightly deviates from the ideal case. Such robustness of the Sally-DA model under detector effects is a clear advantage over Optimal Observables.

Finally, in the last row of \autoref{fig:chi2}, we show the performance on Sample-DB, which includes both detector effects and backgrounds from $e^+e^- \to ZZ$. It is clear that, while the background only contributes 10\% to the total events, failing to incorporate background information results in unacceptably large shifts of central values, as shown by the OO and Sally-D results. The OO results also have uncertainties that are even smaller than the ideal ones (the blue ellipses are smaller). This is another form of bias, suggesting an overestimation of the corresponding precision reach. Instead, Sally-DBA (red) can take account of the background effects and significantly reduce the systematic bias compared to the previous methods, disregarding the effects of backgrounds.     

\begin{figure}
    \centering
    \includegraphics[height=4.5 cm]{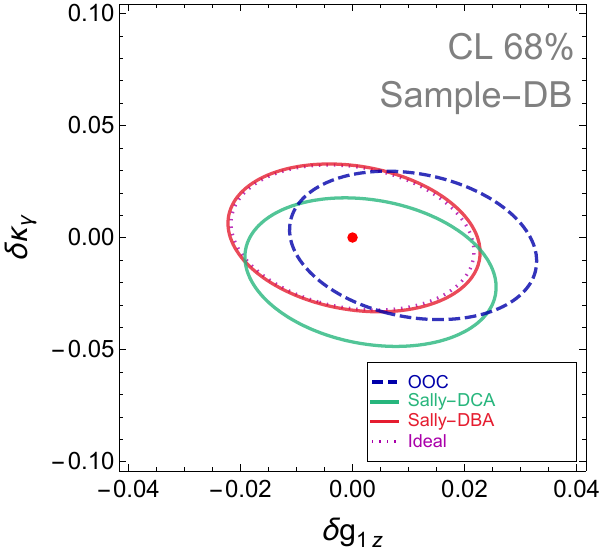}
    \includegraphics[height=4.5 cm]{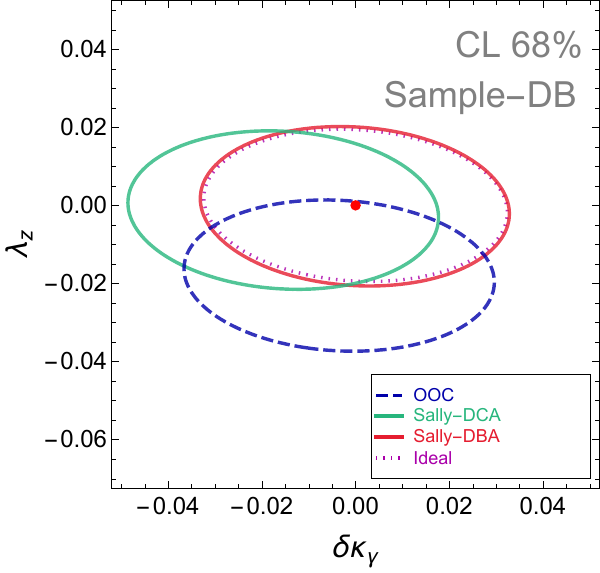}
    \includegraphics[height=4.5 cm]{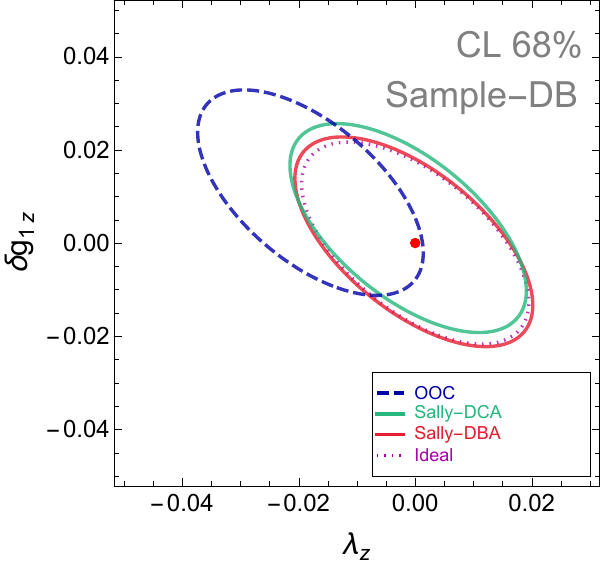}
    \caption{
    Same as the bottom row of \autoref{fig:chi2} (with \textbf{Sample-DB}), but comparing the results of OOC, Sally-DCA and Sally-DBA.
    }
    \label{fig:classifier}
\end{figure}

\begin{figure}
    \centering
    \includegraphics[width=0.8\textwidth]{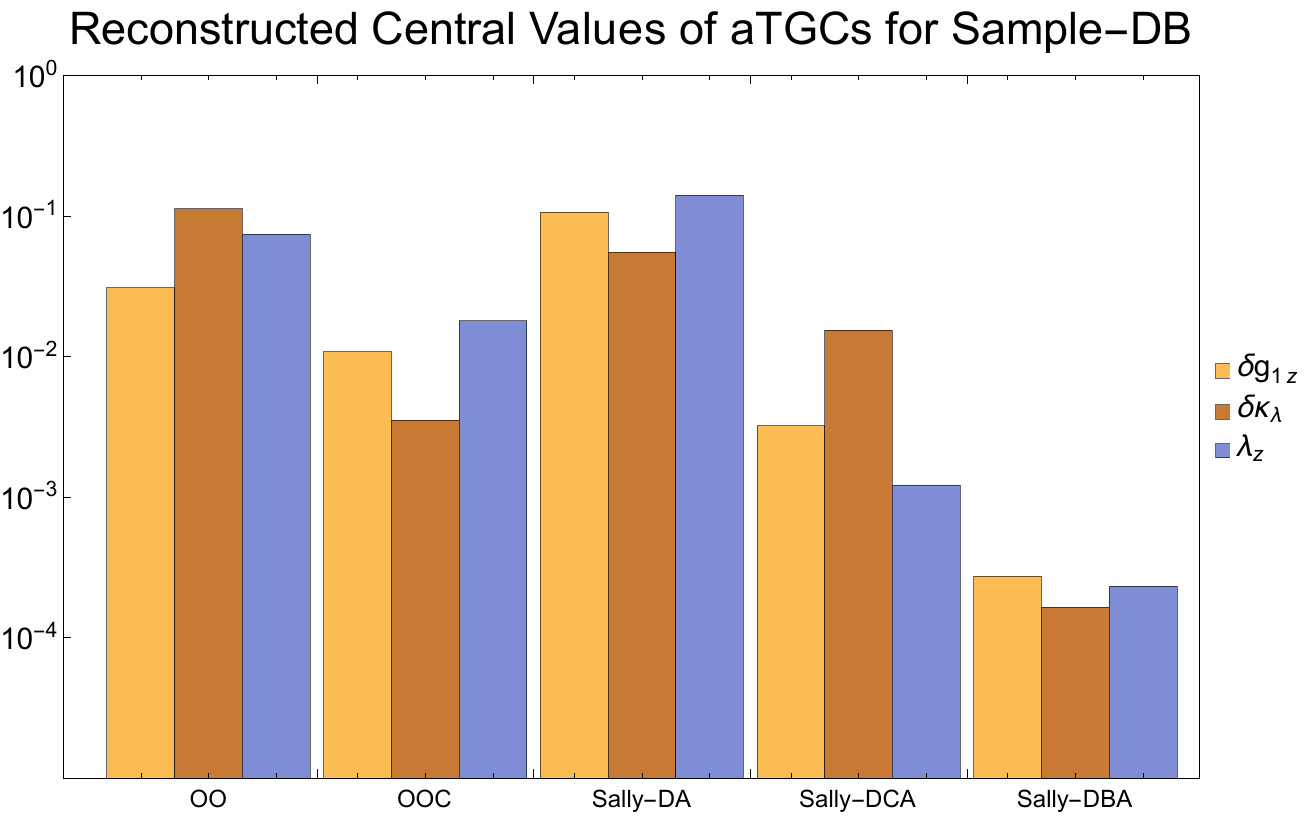}
    \caption{The results in \autoref{tab:bias_method} for Sample-DB presented as bar plots (with only the magnitude of the central values). Note that OO and Sally-DA are generically not applicable to samples with backgrounds, showing significant biases as expected.}
    \label{fig:biasstat}
\end{figure}

In \autoref{fig:classifier}, we further compare the performances of different ML-based strategies on sample-DB. These include OOC, 
Sally-DCA and Sally-DBA. By combining with a classifier filtering out background events, OOC is able to deliver a far better performance than OO. Due to the low background acceptance of the classifier trained, the overall performance of OOC is very similar to the OO method directly applied to Sample-D, which is, however, still unacceptable for the precision reach of future lepton colliders as we discussed previously. In contrast, Sally-DCA is able to handle the impact of both background events and noise from systematic effects, and has a 
slightly better performance than OOC.   
%
It is also notable that Sally-DBA delivers better performance than Sally-DCA with a smaller model thanks to its integrated structure handling all systematic effects simultaneously. Each input event, either signal or background ones, will be assigned with a proper target output $\alpha(z)$. Backgrounds that occasionally pass through the classifier network in Sally-DCA, instead, may sit in a location in the $x$ space where the (signal-event-only) training sample has zero density, giving rise to unregulated $\alpha$ output.
The reconstructed central values of all scenarios are presented in~\autoref{tab:bias_method} for reference and illustrated (for Sample-DB) in~\autoref{fig:biasstat}.  
We also note that the averaging step in Sally-DBA and Sally-DCA (as well as Sally-DA when applied to Sample-D) is important for reaching the desired performance level.  The results for the individual Sally-D, Sally-DB and Sally-DC models are presented in \autoref{app:nn8}.  The performances of the individual models are affected by their modeling and training history, which causes non-negligible systematic error, especially in the central values, as shown in \autoref{fig:chi2nn8}.  The averaging step of the $\alpha$ output reduces the variances picked up by individual models during training.

\begin{table}
\centering
\resizebox{\textwidth}{!}{
\begin{tabular}{|c||c|c|c||c|c|c||c|c|c|}
\hline
         & \multicolumn{3}{c||}{Sample-T}    & \multicolumn{3}{c||}{Sample-D}   & \multicolumn{3}{c|}{Sample-DB}    \\ \hline
         & {$\delta g_{1z}$}        & {$\delta \kappa_\gamma$}      & {$\lambda z$} & {$\delta g_{1z}$}       & {$\delta \kappa_\gamma$}      & {$\lambda z$} & {$\delta g_{1z}$}       & {$\delta \kappa_\gamma$}      & {$\lambda z$} \\ \hline
OO       &  0 & 0  & 0      & {0.009} & {0.0011} & -0.0156     & {0.031} & {-0.11}  & -0.07                 \\ \hline
OOC       &   &  &     &  &  &     & 0.011 &-0.0035  & -0.018              \\ \hline
Sally-T & {0.005} & {-0.0006} & -0.0019  &   &   &    &         &        &     \\ \hline
Sally-DA &     &  &    & {0.0032} & {-0.0029} & -0.00023      & {0.106}    & {0.055} & -0.14    \\ \hline
Sally-DCA       &   &  &     &  &  &     & 0.0032 &-0.015  & -0.0012                 \\ \hline
Sally-DBA       & & &    & &  &      & {0.00027} & {-0.00016}  & -0.00023             \\ \hline
\end{tabular}
}
\caption{The central values of the 3 aTGCs in various scenarios, which provide a measure of the bias of different methods. Note that the ideal central values are zero by construction, and so are the ones for OO on Sample-T due to the calibration (see \autoref{fig:chi2}).}
\label{tab:bias_method}
\end{table}

\begin{figure}
    \centering
    \includegraphics[width=0.9\linewidth]{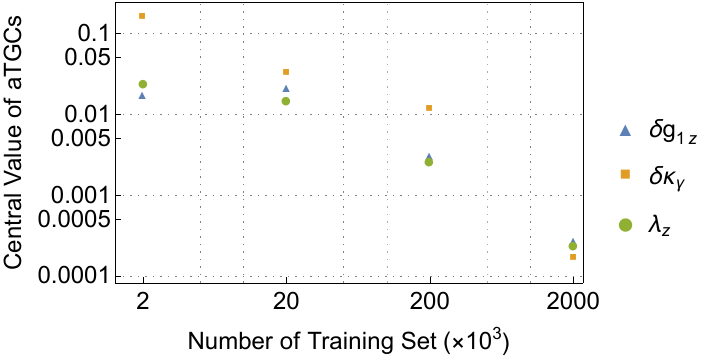}
    \caption{The bias of Sally-DBA (in terms of the central values of aTGCs) trained on different size training set, which shows an decay in the bias for all 3 aTGC parameters.} 
    \label{fig:size}
\end{figure}

Even with Sally-DBA, there is still a bias in the central values of the 3 aTGCs around $\mathcal{O}(10^{-3})$.
This bias is mainly due to limited MC sample size and imperfect training.  
Therefore, with more MC simulation and computing resource (which should be easily achievable in the future), this bias will hopefully be reduced to the desired level for the actual run scenarios of future lepton colliders, which typically has a precision reach of $\sim10^{-3}$--$10^{-4}$ for the 3 aTGCs (so that the bias needs to be below $\mathcal{O}(10^{-4})$).  
%
%
To illustrate this, we show in \autoref{fig:size} the change of the bias in the central values of the 3 aTGCs with Sally-DB where the training sample size is varied from $2\times10^3$ to $2\times10^6$ events.  We can see (up to statistical fluctuations) a clear pattern that the bias decreases with the increase of training statistics, and very roughly scales as $1/\sqrt{N}$ (where $N$ is the total number of training events).  
It is thus reasonable to expect that the bias would continue to decrease as the sample size goes beyond $\sim 10^6$.  Of course, further studies are required to check if the bias could indeed be reduced to the desired level.



\section{Conclusion}
\label{sec:con}

Proposals of various future lepton collider projects have been actively investigated by the particle physics community. While we should remain optimistic that at least one of these proposals will be realized, it is likely that even in the most optimistic scenario, several decades are needed before a sizable amount of data is obtained from any future lepton collider. On the other hand, the field of machine learning, as well as its applications in high energy physics, has been highly active in recent years. With the anticipated developments in both machine-learning algorithms and computing resources, it is foreseeable that machine learning will play a more important role in future collider experiments and become an essential part of many physics analyses. Understanding the physics potential of machine-learning methods is thus an important aspect of future collider studies.

In this paper, we applied machine-learning methods to the SMEFT analysis of the $\eeww$ process at a 240\,GeV lepton collider, and compared their performances with the ones of conventional methods, in particular the Optimal Observables. Our study is based on MC simulations at the detector level.  
We focus on the semi-leptonic channel, $\eeww \to j j \ell \nu$, and consider the $e^+e^- \to ZZ \to j j \ell^+ \ell^-$ background with a missing lepton. 
As a proof of principle, we have shown that optimal observables are subject to unacceptable biases if various systematic effects (such as ISR and detector effects) and backgrounds are not taken into account. On the other hand, the machine-learning version of the optimal observables, SALLY, and its variations, are much more robust under the systematic and background effects. Such methods could be ideal for the SMEFT analyses at future lepton colliders assuming that one has access to accurate MC simulation tools.

The ability to mitigate background effects is essential for future lepton colliders, as the target precision is often very high. Even relatively small SM backgrounds can cause big errors in results. The background events often have profiles similar to signal events but have distinct physics origins. The truth-level description and MC simulation framework of signal and backgrounds could differ a lot accordingly, but it is hard to tell from each other. Facing the challenge, we derived the algorithm to handle general backgrounds. In particular, two approaches are proposed to deliver EFT constraints with proper background mitigation. It is then demonstrated that by either combining a specific classifier network or training the inference network with proper samples, the impact from backgrounds is well under control. The numerical result also suggests that the latter approach with a single network has a moderately higher precision than the case with an extra classifier. 

Limited by time and resources, a few interesting (and challenging) aspects remain to be studied in depth in this work. We hope that future work can address these open questions. One of the most obvious possibilities is to apply the machine-learning methods also to the dileptonic and fully-hadronic channels and evaluate the improvements. In particular, the hadronic channel has a sizable branching ratio but is more challenging due to jet measurement uncertainties and larger backgrounds. Secondly, it is desirable to achieve more realistic MC simulation and model training in future works. Due to the limitation on computing resources, we have used $2\times10^6$ events for training and $5\times10^5$ for validation, which are orders of magnitudes less than the actual sample size at future lepton colliders ($\sim 10^8$). To verify if the observed biases can be indeed decreased to the target level, a much larger signal and background sample size will be necessary. Beyond $\eeww$, one could also apply our methods to other SMEFT analyses, with $\eett$ being an obvious example. It will be interesting to explore different machine-learning techniques and more complex neural network architectures in the hope of finding more efficient and robust methods. As a final remark, we stress that machine learning is most beneficial when theoretical and experimental uncertainties are addressed correctly. In reality, MC simulation may not perfectly match experimental data, which could induce intrinsic (theoretical) uncertainties in the machine learning analysis. Taking account of such uncertainties could be crucial for studies at future lepton colliders. We leave these important directions to future studies.

\subsection*{Acknowledgments}
We thank Henning Bahl, Tianyu He, Claudius Krause, Ying-Ying Li, Andrea Wulzer and Dan Yu for useful discussions. SC and JG are supported by National Natural Science Foundation of China (NSFC) under grant No.~12035008 and No.~12375091.  LL is supported by the DOE grant DE-SC-0010010.   


\appendix

\section{$\chi^2$ in full EFT parameterization}
\label{app:chis}

We hereby provide the $\chi^2$ in the full EFT parameterization (excluding modifications of $m_W$), which corresponds to a 240\,GeV $e^+e^-$ collider with unpolarized beams, using the semi-leptonic channel of $\eeww$, and a total number of $10^4$ signal events (corresponding to $2.2\infb$).  They are provided for three different scenarios for comparison.  Note that the results below actually correspond to $\chi^2-\chi^2_{\rm SM}$ (as in \autoref{eq:chis}), and are zero in the SM limit by construction, while $\chi^2_{\rm min}$ could be smaller than zero if the central values are biased.   

\begin{itemize}

\item Ideal case:
\begin{align}
    \chi ^{2} &=~2401.41\delta \kappa _{\gamma }^{2} +38836.5\delta g_{Z,L}^{e} \delta g_{Z,R}^{e} +119518\delta g_{Z,L}^{e\ 2} -24641.9\delta \kappa _{\gamma } \delta g_{Z,L}^{e} \nonumber\\
     & +220298\delta g_{Z,L}^{e} \delta g_{W}^{l} -54619.2\delta g_{Z,L}^{e} \delta g_{1Z} -41942.3\delta g_{Z,L}^{e} \lambda _{z} +38950.2\delta g_{Z,R}^{e\ 2} \nonumber\\
     & -16465.1\delta \kappa _{\gamma } \delta g_{Z,R}^{e} +30964.6\delta g_{Z,R}^{e} \delta g_{W}^{l} +7846.29\delta g_{Z,R}^{e} \delta g_{1Z} -262.732\delta g_{Z,R}^{e} \lambda _{z} \nonumber\\
     & -21869.9\delta \kappa _{\gamma } \delta g_{W}^{l} +101677\delta g_{W}^{l\ 2} -39096.3\lambda _{z} \delta g_{W}^{l} -51465.1\delta g_{1Z} \delta g_{W}^{l} \nonumber\\
     & +2720.17\delta \kappa _{\gamma } \delta g_{1Z} +2563.75\delta \kappa _{\gamma } \lambda _{z} +11113.9\delta g_{1Z} \lambda _{z} +8192.76\delta g_{1Z}^{2} +9822.05\lambda _{z}^{2} \ .
    \end{align}
\item Sample-D with Sally-DA:
\begin{align}
    \chi ^{2} & =~2371.55\delta \kappa _{\gamma }^{2} +8.17773\delta \kappa _{\gamma } +38835.8\delta g_{Z,L}^{e} \delta g_{Z,R}^{e} +116861\delta g_{Z,L}^{e\ 2}\nonumber \\
     & -24245.7\delta \kappa _{\gamma } \delta g_{Z,L}^{e} +214835\delta g_{Z,L}^{e} \delta g_{W}^{l} -53214.4\delta g_{Z,L}^{e} \delta g_{1Z} -41320.1\delta g_{Z,L}^{e} \lambda _{z}\nonumber \\
     & +134.204\delta g_{Z,L}^{e} +38624.8\delta g_{Z,R}^{e\ 2} -16358.7\delta \kappa _{\gamma } \delta g_{Z,R}^{e} +31316.2\delta g_{Z,R}^{e} \delta g_{W}^{l}\nonumber \\
     & +7693.53\delta g_{Z,R}^{e} \delta g_{1Z} -387.175\delta g_{Z,R}^{e} \lambda _{z} -72.0773\delta g_{Z,R}^{e} -21545.6\delta \kappa _{\gamma } \delta g_{W}^{l}\nonumber \\
     & +125.723\delta g_{W}^{l} +98888.8\delta g_{W}^{l\ 2} -38544.3\lambda _{z} \delta g_{W}^{l} -49941.4\delta g_{1Z} \delta g_{W}^{l}\nonumber \\
     & +2640.43\delta \kappa _{\gamma } \delta g_{1Z} +2574.27\delta \kappa _{\gamma } \lambda _{z} +10920.5\delta g_{1Z} \lambda _{z} +9391.97\delta g_{1Z}^{2}\nonumber \\
     & -51.6591\delta g_{1Z} +9391.97\lambda _{z}^{2} +7.65196\lambda _{z} \ .
    \end{align}
\item Sample-DB with Sally-DBA:
\begin{align}
    \chi ^{2} & = 1376.14\delta \kappa _{\gamma }^{2} -395.321\delta \kappa _{\gamma } +23243.3\delta g_{Z,L}^{e} \delta g_{Z,R}^{e} +95653.3\delta g_{Z,L}^{e\ 2}\nonumber \\
     & -18041\delta \kappa _{\gamma } g_{Z,L}^{e} +171815\delta g_{Z,L}^{e} \delta g_{W}^{l} -45533.2\delta g_{Z,L}^{e} \delta g_{1Z} -34370.6\delta g_{Z,L}^{e} \lambda _{z}\nonumber \\
     & +2752.1\delta g_{Z,L}^{e} +16924.1\delta g_{Z,R}^{e\ 2} -7720.34\delta \kappa _{\gamma } \delta g_{Z,R}^{e} +17404.9\delta g_{Z,R}^{e} \delta g_{W}^{l}\nonumber \\
     & +1619.6\delta g_{Z,R}^{e} \delta g_{1Z} -104.038\delta g_{Z,R}^{e} \lambda _{z} +909.141\delta g_{Z,R}^{e} -15530.9\delta \kappa _{\gamma } \delta g_{W}^{l}\nonumber \\
     & +2731.45\delta g_{W}^{l} +89680.3\delta g_{W}^{l\ 2} -30636\lambda _{z} \delta g_{W}^{l} -41903.2\delta g_{1Z} \delta g_{W}^{l}\nonumber \\
     & +3018.55\delta \kappa _{\gamma } \delta g_{1Z} +2286.83\delta \kappa _{\gamma } \lambda _{z} +9113.05\delta g_{1Z} \lambda _{z} +6264.26\delta g_{1Z}^{2}\nonumber \\
     & -472.895\delta g_{1Z} +5167.5\lambda _{z}^{2} -332.831\lambda _{z} \ 
    \end{align}

\end{itemize}

\section{Results of individual neutral network models}
\label{app:nn8}

\begin{figure}
    \centering
    \includegraphics[height=4.5 cm]{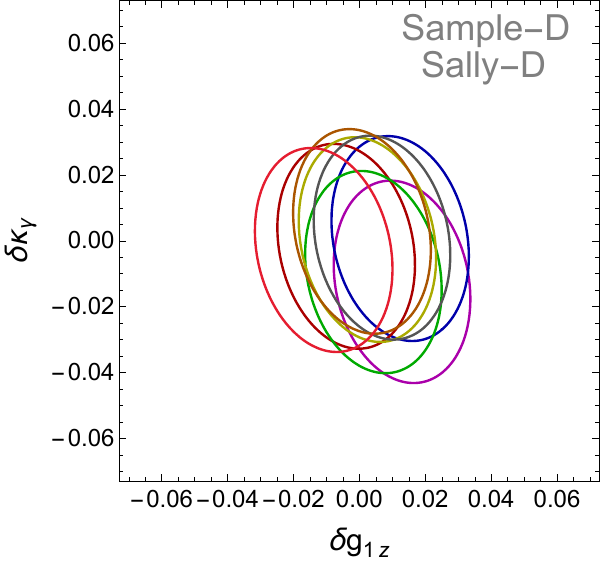}
    \includegraphics[height=4.5 cm]{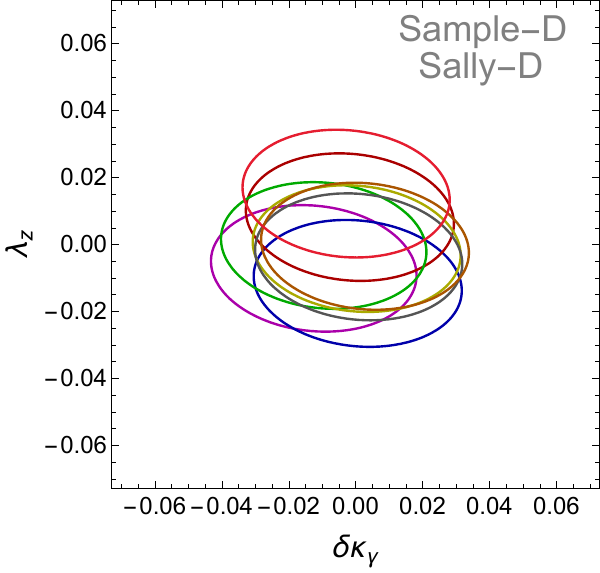}
    \includegraphics[height=4.5 cm]{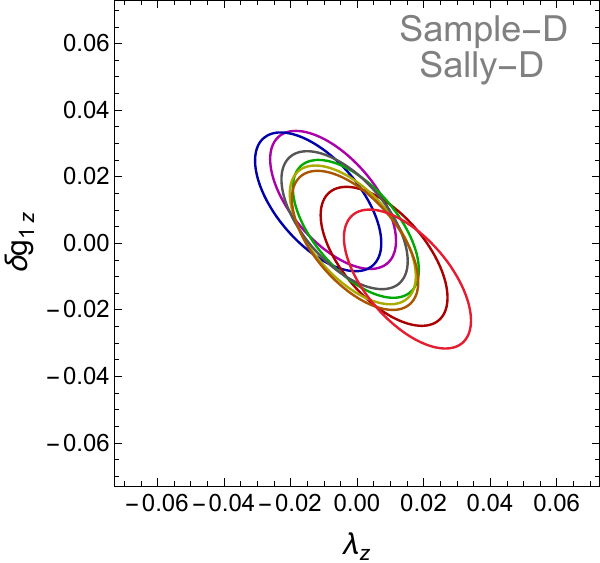}\\
    \includegraphics[height=4.5 cm]{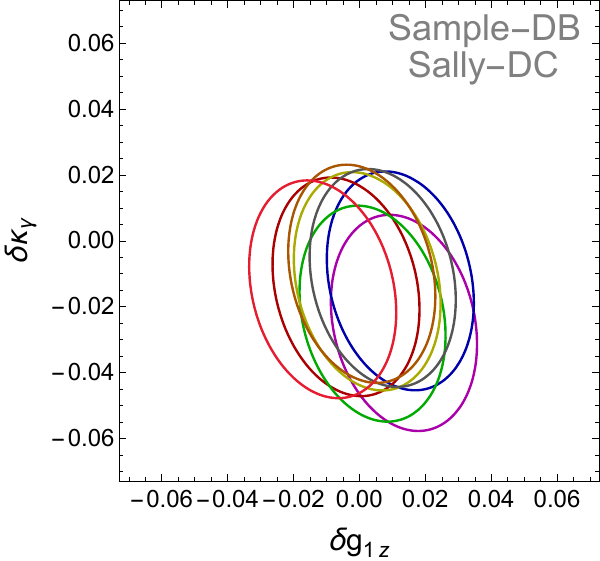}
    \includegraphics[height=4.5 cm]{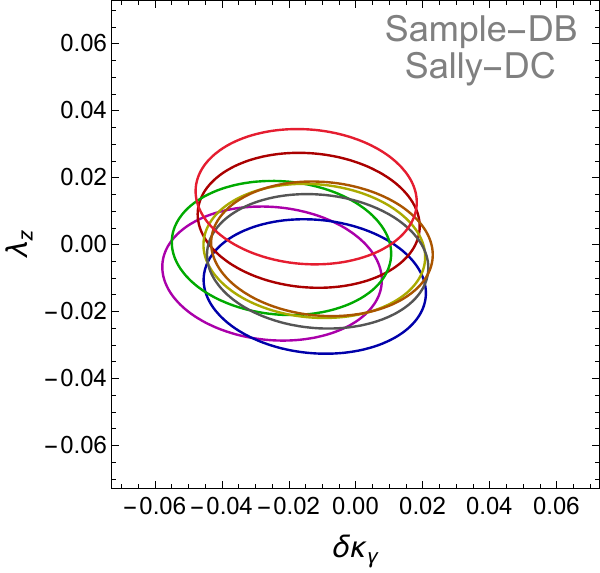}
    \includegraphics[height=4.5 cm]{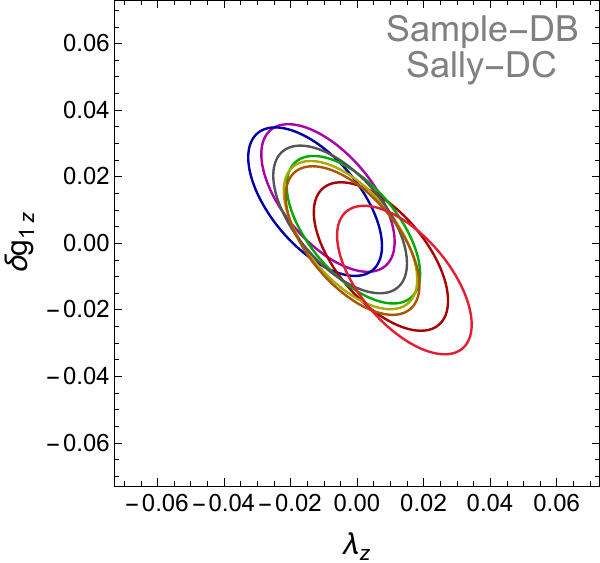}\\
    \includegraphics[height=4.5 cm]{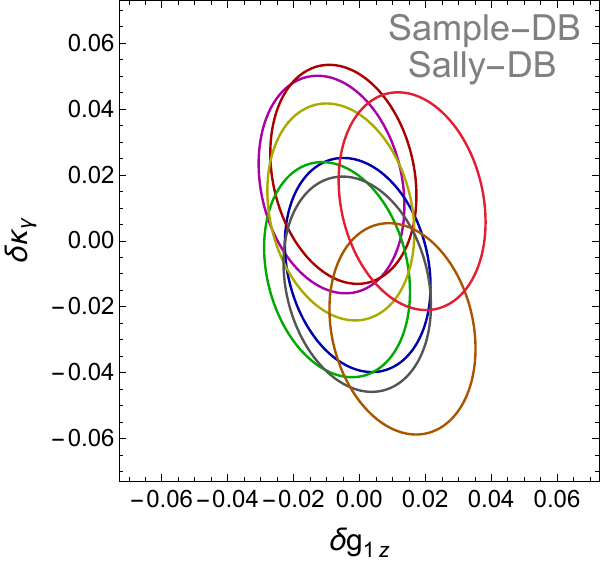}
    \includegraphics[height=4.5 cm]{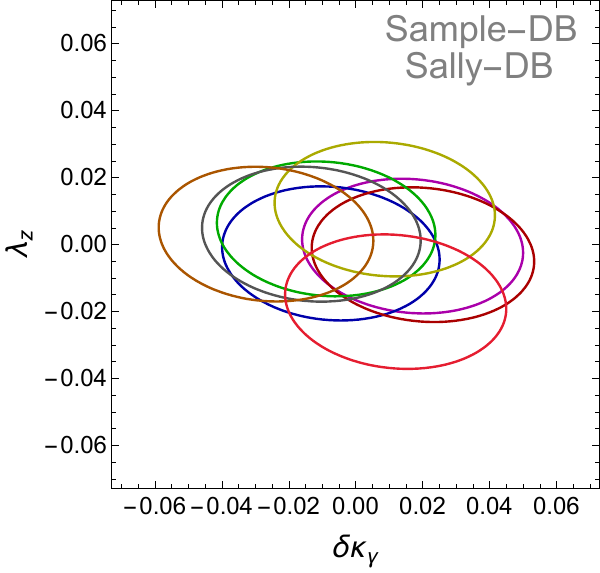}
    \includegraphics[height=4.5 cm]{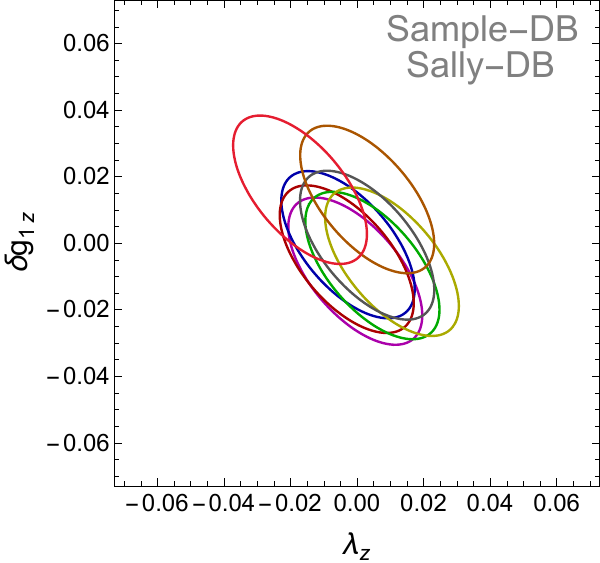}\\
    \caption{
    The 68\%CL contours as in \autoref{fig:chi2} but displaying the results of the 8 (un-averaged) neural network models separately.  {\bf Top:} Sally-D applied to Sample-D.  {\bf Middle:} Sally-DC applied to Sample-DB.  {\bf Bottom:} Sally-DB applied to Sample-DB. 
    }
    \label{fig:chi2nn8}
\end{figure}

As mentioned in \autoref{sec:simulation} and \autoref{sec:result}, for each one of Sally-DA, Sally-DCA and Sally-DBA, we train 8 neural network models and take the average of the reconstructed $\hat{\alpha}$ of the 8 models to produce the final result.  This averaging step is important for reducing the systematic error of individual models due to imperfect training.  To illustrate this, we show in \autoref{fig:chi2nn8} the limits on the 3 aTGCs obtained from the 8 individual models of Sally-DA (top row), Sally-DC (middle row) and Sally-DB (bottom row).  In each case, it is clear from \autoref{fig:chi2nn8} that the results of the 8 individual models have a significant spread.  In particular, the reconstructed central values could go as far as $\sim \pm 0.03$ for $\delta \kappa_\gamma$, clearly unacceptable in our case.  However, in most cases the spread is more or less symmetrical in the positive and negative directions, so the averaging process is very helpful in reducing the overall error.  

\bibliographystyle{JHEP}
\bibliography{gen}

\providecommand{\href}[2]{#2}\begingroup\raggedright\begin{thebibliography}{10}

\bibitem{Narain:2022qud}
M.~Narain et~al., \emph{{The Future of US Particle Physics - The Snowmass 2021
  Energy Frontier Report}},  \href{https://arxiv.org/abs/2211.11084}{{\ttfamily
  2211.11084}}.

\bibitem{CEPCPhysicsStudyGroup:2022uwl}
{\scshape CEPC Physics Study Group} collaboration, \emph{{The Physics potential
  of the CEPC. Prepared for the US Snowmass Community Planning Exercise
  (Snowmass 2021)}},  in \emph{{2022 Snowmass Summer Study}}, 5, 2022,
  \href{https://arxiv.org/abs/2205.08553}{{\ttfamily 2205.08553}}.

\bibitem{Bernardi:2022hny}
G.~Bernardi et~al., \emph{{The Future Circular Collider: a Summary for the US
  2021 Snowmass Process}},  \href{https://arxiv.org/abs/2203.06520}{{\ttfamily
  2203.06520}}.

\bibitem{ILCInternationalDevelopmentTeam:2022izu}
{\scshape ILC International Development Team} collaboration, \emph{{The
  International Linear Collider: Report to Snowmass 2021}},
  \href{https://arxiv.org/abs/2203.07622}{{\ttfamily 2203.07622}}.

\bibitem{CLICdp:2018cto}
{\scshape CLICdp, CLIC} collaboration, \emph{{The Compact Linear Collider
  (CLIC) - 2018 Summary Report}},
  \href{https://arxiv.org/abs/1812.06018}{{\ttfamily 1812.06018}}.

\bibitem{Bai:2021rdg}
M.~Bai et~al., \emph{{C$^3$: A ''Cool'' Route to the Higgs Boson and Beyond}},
  in \emph{{2022 Snowmass Summer Study}}, 10, 2021,
  \href{https://arxiv.org/abs/2110.15800}{{\ttfamily 2110.15800}}.

\bibitem{Aime:2022flm}
C.~Aime et~al., \emph{{Muon Collider Physics Summary}},
  \href{https://arxiv.org/abs/2203.07256}{{\ttfamily 2203.07256}}.

\bibitem{Accettura:2023ked}
C.~Accettura et~al., \emph{{Towards a muon collider}},
  \href{https://doi.org/10.1140/epjc/s10052-023-11889-x}{\emph{Eur. Phys. J. C}
  {\bfseries 83} (2023) 864}
  [\href{https://arxiv.org/abs/2303.08533}{{\ttfamily 2303.08533}}].

\bibitem{deBlas:2019rxi}
J.~de~Blas et~al., \emph{{Higgs Boson Studies at Future Particle Colliders}},
  \href{https://doi.org/10.1007/JHEP01(2020)139}{\emph{JHEP} {\bfseries 01}
  (2020) 139} [\href{https://arxiv.org/abs/1905.03764}{{\ttfamily
  1905.03764}}].

\bibitem{deBlas:2022aow}
J.~de~Blas, J.~Gu and Z.~Liu, \emph{{Higgs boson precision measurements at a
  125~GeV muon collider}},
  \href{https://doi.org/10.1103/PhysRevD.106.073007}{\emph{Phys. Rev. D}
  {\bfseries 106} (2022) 073007}
  [\href{https://arxiv.org/abs/2203.04324}{{\ttfamily 2203.04324}}].

\bibitem{Forslund:2022xjq}
M.~Forslund and P.~Meade, \emph{{High precision higgs from high energy muon
  colliders}}, \href{https://doi.org/10.1007/JHEP08(2022)185}{\emph{JHEP}
  {\bfseries 08} (2022) 185}
  [\href{https://arxiv.org/abs/2203.09425}{{\ttfamily 2203.09425}}].

\bibitem{Falkowski:2015jaa}
A.~Falkowski, M.~Gonzalez-Alonso, A.~Greljo and D.~Marzocca, \emph{{Global
  constraints on anomalous triple gauge couplings in effective field theory
  approach}}, \href{https://doi.org/10.1103/PhysRevLett.116.011801}{\emph{Phys.
  Rev. Lett.} {\bfseries 116} (2016) 011801}
  [\href{https://arxiv.org/abs/1508.00581}{{\ttfamily 1508.00581}}].

\bibitem{Durieux:2017rsg}
G.~Durieux, C.~Grojean, J.~Gu and K.~Wang, \emph{{The leptonic future of the
  Higgs}}, \href{https://doi.org/10.1007/JHEP09(2017)014}{\emph{JHEP}
  {\bfseries 09} (2017) 014}
  [\href{https://arxiv.org/abs/1704.02333}{{\ttfamily 1704.02333}}].

\bibitem{Barklow:2017suo}
T.~Barklow, K.~Fujii, S.~Jung, R.~Karl, J.~List, T.~Ogawa et~al.,
  \emph{{Improved Formalism for Precision Higgs Coupling Fits}},
  \href{https://doi.org/10.1103/PhysRevD.97.053003}{\emph{Phys. Rev. D}
  {\bfseries 97} (2018) 053003}
  [\href{https://arxiv.org/abs/1708.08912}{{\ttfamily 1708.08912}}].

\bibitem{Durieux:2018tev}
G.~Durieux, M.~Perell\'o, M.~Vos and C.~Zhang, \emph{{Global and optimal probes
  for the top-quark effective field theory at future lepton colliders}},
  \href{https://doi.org/10.1007/JHEP10(2018)168}{\emph{JHEP} {\bfseries 10}
  (2018) 168} [\href{https://arxiv.org/abs/1807.02121}{{\ttfamily
  1807.02121}}].

\bibitem{Durieux:2018ggn}
G.~Durieux, J.~Gu, E.~Vryonidou and C.~Zhang, \emph{{Probing top-quark
  couplings indirectly at Higgs factories}},
  \href{https://doi.org/10.1088/1674-1137/42/12/123107}{\emph{Chin. Phys. C}
  {\bfseries 42} (2018) 123107}
  [\href{https://arxiv.org/abs/1809.03520}{{\ttfamily 1809.03520}}].

\bibitem{Ellis:2018gqa}
J.~Ellis, C.~W. Murphy, V.~Sanz and T.~You, \emph{{Updated Global SMEFT Fit to
  Higgs, Diboson and Electroweak Data}},
  \href{https://doi.org/10.1007/JHEP06(2018)146}{\emph{JHEP} {\bfseries 06}
  (2018) 146} [\href{https://arxiv.org/abs/1803.03252}{{\ttfamily
  1803.03252}}].

\bibitem{DeBlas:2019qco}
J.~De~Blas, G.~Durieux, C.~Grojean, J.~Gu and A.~Paul, \emph{{On the future of
  Higgs, electroweak and diboson measurements at lepton colliders}},
  \href{https://doi.org/10.1007/JHEP12(2019)117}{\emph{JHEP} {\bfseries 12}
  (2019) 117} [\href{https://arxiv.org/abs/1907.04311}{{\ttfamily
  1907.04311}}].

\bibitem{Durieux:2019rbz}
G.~Durieux, A.~Irles, V.~Miralles, A.~Pe\~nuelas, R.~P\"oschl, M.~Perell\'o
  et~al., \emph{{The electro-weak couplings of the top and bottom quarks
  \textemdash{} Global fit and future prospects}},
  \href{https://doi.org/10.1007/JHEP12(2019)098}{\emph{JHEP} {\bfseries 12}
  (2019) 98} [\href{https://arxiv.org/abs/1907.10619}{{\ttfamily 1907.10619}}].

\bibitem{Ellis:2020unq}
J.~Ellis, M.~Madigan, K.~Mimasu, V.~Sanz and T.~You, \emph{{Top, Higgs, Diboson
  and Electroweak Fit to the Standard Model Effective Field Theory}},
  \href{https://doi.org/10.1007/JHEP04(2021)279}{\emph{JHEP} {\bfseries 04}
  (2021) 279} [\href{https://arxiv.org/abs/2012.02779}{{\ttfamily
  2012.02779}}].

\bibitem{Liu:2022vgo}
Y.~Liu, Y.~Wang, C.~Zhang, L.~Zhang and J.~Gu, \emph{{Probing top-quark
  operators with precision electroweak measurements}},
  \href{https://doi.org/10.1088/1674-1137/ac82e1}{\emph{Chin. Phys. C}
  {\bfseries 46} (2022) 113105}
  [\href{https://arxiv.org/abs/2205.05655}{{\ttfamily 2205.05655}}].

\bibitem{deBlas:2022ofj}
J.~de~Blas, Y.~Du, C.~Grojean, J.~Gu, V.~Miralles, M.~E. Peskin et~al.,
  \emph{{Global SMEFT Fits at Future Colliders}},  in \emph{{2022 Snowmass
  Summer Study}}, 6, 2022, \href{https://arxiv.org/abs/2206.08326}{{\ttfamily
  2206.08326}}.

\bibitem{Ethier:2021bye}
{\scshape SMEFiT} collaboration, \emph{{Combined SMEFT interpretation of Higgs,
  diboson, and top quark data from the LHC}},
  \href{https://doi.org/10.1007/JHEP11(2021)089}{\emph{JHEP} {\bfseries 11}
  (2021) 089} [\href{https://arxiv.org/abs/2105.00006}{{\ttfamily
  2105.00006}}].

\bibitem{Brivio:2022hrb}
I.~Brivio, S.~Bruggisser, N.~Elmer, E.~Geoffray, M.~Luchmann and T.~Plehn,
  \emph{{To Profile or To Marginalize -- A SMEFT Case Study}},
  \href{https://arxiv.org/abs/2208.08454}{{\ttfamily 2208.08454}}.

\bibitem{Bartocci:2023nvp}
R.~Bartocci, A.~Biek\"otter and T.~Hurth, \emph{{A global analysis of the SMEFT
  under the minimal MFV assumption}},
  \href{https://arxiv.org/abs/2311.04963}{{\ttfamily 2311.04963}}.

\bibitem{Allwicher:2023shc}
L.~Allwicher, C.~Cornella, G.~Isidori and B.~A. Stefanek, \emph{{New Physics in
  the Third Generation: A Comprehensive SMEFT Analysis and Future Prospects}},
  \href{https://arxiv.org/abs/2311.00020}{{\ttfamily 2311.00020}}.

\bibitem{Wen:2023xxc}
X.-K. Wen, B.~Yan, Z.~Yu and C.~P. Yuan, \emph{{Single Transverse Spin
  Asymmetry as a New Probe of Standard-Model-Effective-Field-Theory Dipole
  Operators}},
  \href{https://doi.org/10.1103/PhysRevLett.131.241801}{\emph{Phys. Rev. Lett.}
  {\bfseries 131} (2023) 241801}
  [\href{https://arxiv.org/abs/2307.05236}{{\ttfamily 2307.05236}}].

\bibitem{Marchesini:2011aka}
I.~Marchesini, \emph{{Triple gauge couplings and polarization at the ILC and
  leakage in a highly granular calorimeter}}, Ph.D. thesis, Hamburg U., 2011.

\bibitem{Bian:2015zha}
L.~Bian, J.~Shu and Y.~Zhang, \emph{{Prospects for Triple Gauge Coupling
  Measurements at Future Lepton Colliders and the 14 TeV LHC}},
  \href{https://doi.org/10.1007/JHEP09(2015)206}{\emph{JHEP} {\bfseries 09}
  (2015) 206} [\href{https://arxiv.org/abs/1507.02238}{{\ttfamily
  1507.02238}}].

\bibitem{Grojean:2018dqj}
C.~Grojean, M.~Montull and M.~Riembau, \emph{{Diboson at the LHC vs LEP}},
  \href{https://doi.org/10.1007/JHEP03(2019)020}{\emph{JHEP} {\bfseries 03}
  (2019) 020} [\href{https://arxiv.org/abs/1810.05149}{{\ttfamily
  1810.05149}}].

\bibitem{Subba:2022czw}
A.~Subba and R.~K. Singh, \emph{{Role of polarizations and spin-spin
  correlations of W's in e-e+$\to$W-W+ at s=250\,\,GeV to probe anomalous
  W-W+Z/$\gamma$ couplings}},
  \href{https://doi.org/10.1103/PhysRevD.107.073004}{\emph{Phys. Rev. D}
  {\bfseries 107} (2023) 073004}
  [\href{https://arxiv.org/abs/2212.12973}{{\ttfamily 2212.12973}}].

\bibitem{Subba:2023rpm}
A.~Subba and R.~K. Singh, \emph{{Study of anomalous $W^-W^+\gamma /Z$ couplings
  using polarizations and spin correlations in $e^-e^+\rightarrow W^-W^+$ with
  polarized beams}},
  \href{https://doi.org/10.1140/epjc/s10052-023-12292-2}{\emph{Eur. Phys. J. C}
  {\bfseries 83} (2023) 1119}
  [\href{https://arxiv.org/abs/2305.15106}{{\ttfamily 2305.15106}}].

\bibitem{Diehl:1993br}
M.~Diehl and O.~Nachtmann, \emph{{Optimal observables for the measurement of
  three gauge boson couplings in e+ e- ---\ensuremath{>} W+ W-}},
  \href{https://doi.org/10.1007/BF01555899}{\emph{Z. Phys. C} {\bfseries 62}
  (1994) 397}.

\bibitem{Contino:2016jqw}
R.~Contino, A.~Falkowski, F.~Goertz, C.~Grojean and F.~Riva, \emph{{On the
  Validity of the Effective Field Theory Approach to SM Precision Tests}},
  \href{https://doi.org/10.1007/JHEP07(2016)144}{\emph{JHEP} {\bfseries 07}
  (2016) 144} [\href{https://arxiv.org/abs/1604.06444}{{\ttfamily
  1604.06444}}].

\bibitem{Alte:2017pme}
S.~Alte, M.~K\"onig and W.~Shepherd, \emph{{Consistent Searches for SMEFT
  Effects in Non-Resonant Dijet Events}},
  \href{https://doi.org/10.1007/JHEP01(2018)094}{\emph{JHEP} {\bfseries 01}
  (2018) 094} [\href{https://arxiv.org/abs/1711.07484}{{\ttfamily
  1711.07484}}].

\bibitem{Brehmer:2018kdj}
J.~Brehmer, K.~Cranmer, G.~Louppe and J.~Pavez, \emph{{Constraining Effective
  Field Theories with Machine Learning}},
  \href{https://doi.org/10.1103/PhysRevLett.121.111801}{\emph{Phys. Rev. Lett.}
  {\bfseries 121} (2018) 111801}
  [\href{https://arxiv.org/abs/1805.00013}{{\ttfamily 1805.00013}}].

\bibitem{Brehmer:2018eca}
J.~Brehmer, K.~Cranmer, G.~Louppe and J.~Pavez, \emph{{A Guide to Constraining
  Effective Field Theories with Machine Learning}},
  \href{https://doi.org/10.1103/PhysRevD.98.052004}{\emph{Phys. Rev. D}
  {\bfseries 98} (2018) 052004}
  [\href{https://arxiv.org/abs/1805.00020}{{\ttfamily 1805.00020}}].

\bibitem{Brehmer:2018hga}
J.~Brehmer, G.~Louppe, J.~Pavez and K.~Cranmer, \emph{{Mining gold from
  implicit models to improve likelihood-free inference}},
  \href{https://doi.org/10.1073/pnas.1915980117}{\emph{Proc. Nat. Acad. Sci.}
  {\bfseries 117} (2020) 5242}
  [\href{https://arxiv.org/abs/1805.12244}{{\ttfamily 1805.12244}}].

\bibitem{Brehmer:2019xox}
J.~Brehmer, F.~Kling, I.~Espejo and K.~Cranmer, \emph{{MadMiner: Machine
  learning-based inference for particle physics}},
  \href{https://doi.org/10.1007/s41781-020-0035-2}{\emph{Comput. Softw. Big
  Sci.} {\bfseries 4} (2020) 3}
  [\href{https://arxiv.org/abs/1907.10621}{{\ttfamily 1907.10621}}].

\bibitem{DAgnolo:2019vbw}
R.~T. D'Agnolo, G.~Grosso, M.~Pierini, A.~Wulzer and M.~Zanetti,
  \emph{{Learning multivariate new physics}},
  \href{https://doi.org/10.1140/epjc/s10052-021-08853-y}{\emph{Eur. Phys. J. C}
  {\bfseries 81} (2021) 89} [\href{https://arxiv.org/abs/1912.12155}{{\ttfamily
  1912.12155}}].

\bibitem{Chen:2020mev}
S.~Chen, A.~Glioti, G.~Panico and A.~Wulzer, \emph{{Parametrized classifiers
  for optimal EFT sensitivity}},
  \href{https://doi.org/10.1007/JHEP05(2021)247}{\emph{JHEP} {\bfseries 05}
  (2021) 247} [\href{https://arxiv.org/abs/2007.10356}{{\ttfamily
  2007.10356}}].

\bibitem{Chen:2023ind}
S.~Chen, A.~Glioti, G.~Panico and A.~Wulzer, \emph{{Boosting likelihood
  learning with event reweighting}},
  \href{https://arxiv.org/abs/2308.05704}{{\ttfamily 2308.05704}}.

\bibitem{Brehmer:2019gmn}
J.~Brehmer, S.~Dawson, S.~Homiller, F.~Kling and T.~Plehn, \emph{{Benchmarking
  simplified template cross sections in $WH$ production}},
  \href{https://doi.org/10.1007/JHEP11(2019)034}{\emph{JHEP} {\bfseries 11}
  (2019) 034} [\href{https://arxiv.org/abs/1908.06980}{{\ttfamily
  1908.06980}}].

\bibitem{Butter:2021rvz}
A.~Butter, T.~Plehn, N.~Soybelman and J.~Brehmer, \emph{{Back to the Formula --
  LHC Edition}},  \href{https://arxiv.org/abs/2109.10414}{{\ttfamily
  2109.10414}}.

\bibitem{Chatterjee:2021nms}
S.~Chatterjee, N.~Frohner, L.~Lechner, R.~Sch\"ofbeck and D.~Schwarz,
  \emph{{Tree boosting for learning EFT parameters}},
  \href{https://doi.org/10.1016/j.cpc.2022.108385}{\emph{Comput. Phys. Commun.}
  {\bfseries 277} (2022) 108385}
  [\href{https://arxiv.org/abs/2107.10859}{{\ttfamily 2107.10859}}].

\bibitem{GomezAmbrosio:2022mpm}
R.~Gomez~Ambrosio, J.~ter Hoeve, M.~Madigan, J.~Rojo and V.~Sanz,
  \emph{{Unbinned multivariate observables for global SMEFT analyses from
  machine learning}},
  \href{https://doi.org/10.1007/JHEP03(2023)033}{\emph{JHEP} {\bfseries 03}
  (2023) 033} [\href{https://arxiv.org/abs/2211.02058}{{\ttfamily
  2211.02058}}].

\bibitem{Arganda:2022qzy}
E.~Arganda, X.~Marcano, V.~M. Lozano, A.~D. Medina, A.~D. Perez, M.~Szewc
  et~al., \emph{{A method for approximating optimal statistical significances
  with machine-learned likelihoods}},
  \href{https://doi.org/10.1140/epjc/s10052-022-10944-3}{\emph{Eur. Phys. J. C}
  {\bfseries 82} (2022) 993}
  [\href{https://arxiv.org/abs/2205.05952}{{\ttfamily 2205.05952}}].

\bibitem{Chatterjee:2022oco}
S.~Chatterjee, S.~Rohshap, R.~Sch\"ofbeck and D.~Schwarz, \emph{{Learning the
  EFT likelihood with tree boosting}},
  \href{https://arxiv.org/abs/2205.12976}{{\ttfamily 2205.12976}}.

\bibitem{Grojean:2022mef}
C.~Grojean, A.~Paul, Z.~Qian and I.~Str\"umke, \emph{{Lessons on interpretable
  machine learning from particle physics}},
  \href{https://doi.org/10.1038/s42254-022-00456-0}{\emph{Nature Rev. Phys.}
  {\bfseries 4} (2022) 284} [\href{https://arxiv.org/abs/2203.08021}{{\ttfamily
  2203.08021}}].

\bibitem{Alasfar:2022vqw}
L.~Alasfar, R.~Gr\"ober, C.~Grojean, A.~Paul and Z.~Qian, \emph{{Machine
  learning the trilinear and light-quark Yukawa couplings from Higgs pair
  kinematic shapes}},
  \href{https://doi.org/10.1007/JHEP11(2022)045}{\emph{JHEP} {\bfseries 11}
  (2022) 045} [\href{https://arxiv.org/abs/2207.04157}{{\ttfamily
  2207.04157}}].

\bibitem{Letizia:2022xbe}
M.~Letizia, G.~Losapio, M.~Rando, G.~Grosso, A.~Wulzer, M.~Pierini et~al.,
  \emph{{Learning new physics efficiently with nonparametric methods}},
  \href{https://doi.org/10.1140/epjc/s10052-022-10830-y}{\emph{Eur. Phys. J. C}
  {\bfseries 82} (2022) 879}
  [\href{https://arxiv.org/abs/2204.02317}{{\ttfamily 2204.02317}}].

\bibitem{Li:2020vav}
L.~Li, Y.-Y. Li, T.~Liu and S.-J. Xu, \emph{{Learning physics at future $e^-
  e^+$ colliders with machine}},
  \href{https://doi.org/10.1007/JHEP10(2020)018}{\emph{JHEP} {\bfseries 10}
  (2020) 018} [\href{https://arxiv.org/abs/2004.15013}{{\ttfamily
  2004.15013}}].

\bibitem{Yang:2021kyy}
J.-C. Yang, Y.-C. Guo and L.-H. Cai, \emph{{Using a nested anomaly detection
  machine learning algorithm to study the neutral triple gauge couplings at an
  e+e- collider}},
  \href{https://doi.org/10.1016/j.nuclphysb.2022.115735}{\emph{Nucl. Phys. B}
  {\bfseries 977} (2022) 115735}
  [\href{https://arxiv.org/abs/2111.10543}{{\ttfamily 2111.10543}}].

\bibitem{Yang:2022fhw}
J.-C. Yang, X.-Y. Han, Z.-B. Qin, T.~Li and Y.-C. Guo, \emph{{Measuring the
  anomalous quartic gauge couplings in the $W^+W^- \to W^+W^-$ process at muon
  collider using artificial neural networks}},
  \href{https://doi.org/10.1007/JHEP09(2022)074}{\emph{JHEP} {\bfseries 09}
  (2022) 074} [\href{https://arxiv.org/abs/2204.10034}{{\ttfamily
  2204.10034}}].

\bibitem{Dong:2023nir}
Y.-F. Dong, Y.-C. Mao, i.-C. Yang and J.-C. Yang, \emph{{Searching for
  anomalous quartic gauge couplings at muon colliders using principal component
  analysis}}, \href{https://doi.org/10.1140/epjc/s10052-023-11719-0}{\emph{Eur.
  Phys. J. C} {\bfseries 83} (2023) 555}
  [\href{https://arxiv.org/abs/2304.01505}{{\ttfamily 2304.01505}}].

\bibitem{Metodiev:2017vrx}
E.~M. Metodiev, B.~Nachman and J.~Thaler, \emph{{Classification without labels:
  Learning from mixed samples in high energy physics}},
  \href{https://doi.org/10.1007/JHEP10(2017)174}{\emph{JHEP} {\bfseries 10}
  (2017) 174} [\href{https://arxiv.org/abs/1708.02949}{{\ttfamily
  1708.02949}}].

\bibitem{Nachman:2021yvi}
B.~Nachman and J.~Thaler, \emph{{Learning from many collider events at once}},
  \href{https://doi.org/10.1103/PhysRevD.103.116013}{\emph{Phys. Rev. D}
  {\bfseries 103} (2021) 116013}
  [\href{https://arxiv.org/abs/2101.07263}{{\ttfamily 2101.07263}}].

\bibitem{Gambhir:2022dut}
R.~Gambhir, B.~Nachman and J.~Thaler, \emph{{Bias and priors in machine
  learning calibrations for high energy physics}},
  \href{https://doi.org/10.1103/PhysRevD.106.036011}{\emph{Phys. Rev. D}
  {\bfseries 106} (2022) 036011}
  [\href{https://arxiv.org/abs/2205.05084}{{\ttfamily 2205.05084}}].

\bibitem{Gambhir:2022gua}
R.~Gambhir, B.~Nachman and J.~Thaler, \emph{{Learning Uncertainties the
  Frequentist Way: Calibration and Correlation in High Energy Physics}},
  \href{https://doi.org/10.1103/PhysRevLett.129.082001}{\emph{Phys. Rev. Lett.}
  {\bfseries 129} (2022) 082001}
  [\href{https://arxiv.org/abs/2205.03413}{{\ttfamily 2205.03413}}].

\bibitem{Feickert:2021ajf}
M.~Feickert and B.~Nachman, \emph{{A Living Review of Machine Learning for
  Particle Physics}},  \href{https://arxiv.org/abs/2102.02770}{{\ttfamily
  2102.02770}}.

\bibitem{dAgnolo:2021aun}
R.~T. d'Agnolo, G.~Grosso, M.~Pierini, A.~Wulzer and M.~Zanetti,
  \emph{{Learning new physics from an imperfect machine}},
  \href{https://doi.org/10.1140/epjc/s10052-022-10226-y}{\emph{Eur. Phys. J. C}
  {\bfseries 82} (2022) 275}
  [\href{https://arxiv.org/abs/2111.13633}{{\ttfamily 2111.13633}}].

\bibitem{Arganda:2022zbs}
E.~Arganda, A.~D. Perez, M.~de~los Rios and R.~M. Sand\'a~Seoane,
  \emph{{Machine-learned exclusion limits without binning}},
  \href{https://doi.org/10.1140/epjc/s10052-023-12314-z}{\emph{Eur. Phys. J. C}
  {\bfseries 83} (2023) 1158}
  [\href{https://arxiv.org/abs/2211.04806}{{\ttfamily 2211.04806}}].

\bibitem{Arganda:2023qni}
E.~Arganda, D.~A. D\'\i{}az, A.~D. Perez, R.~M. Sand\'a~Seoane and A.~Szynkman,
  \emph{{LHC Study of Third-Generation Scalar Leptoquarks with Machine-Learned
  Likelihoods}},  \href{https://arxiv.org/abs/2309.05407}{{\ttfamily
  2309.05407}}.

\bibitem{Guest:2018yhq}
D.~Guest, K.~Cranmer and D.~Whiteson, \emph{{Deep Learning and its Application
  to LHC Physics}},
  \href{https://doi.org/10.1146/annurev-nucl-101917-021019}{\emph{Ann. Rev.
  Nucl. Part. Sci.} {\bfseries 68} (2018) 161}
  [\href{https://arxiv.org/abs/1806.11484}{{\ttfamily 1806.11484}}].

\bibitem{Carleo:2019ptp}
G.~Carleo, I.~Cirac, K.~Cranmer, L.~Daudet, M.~Schuld, N.~Tishby et~al.,
  \emph{{Machine learning and the physical sciences}},
  \href{https://doi.org/10.1103/RevModPhys.91.045002}{\emph{Rev. Mod. Phys.}
  {\bfseries 91} (2019) 045002}
  [\href{https://arxiv.org/abs/1903.10563}{{\ttfamily 1903.10563}}].

\bibitem{Cranmer:2019eaq}
K.~Cranmer, J.~Brehmer and G.~Louppe, \emph{{The frontier of simulation-based
  inference}}, \href{https://doi.org/10.1073/pnas.1912789117}{\emph{Proc. Nat.
  Acad. Sci.} {\bfseries 117} (2020) 30055}
  [\href{https://arxiv.org/abs/1911.01429}{{\ttfamily 1911.01429}}].

\bibitem{Karagiorgi:2021ngt}
G.~Karagiorgi, G.~Kasieczka, S.~Kravitz, B.~Nachman and D.~Shih, \emph{{Machine
  Learning in the Search for New Fundamental Physics}},
  \href{https://arxiv.org/abs/2112.03769}{{\ttfamily 2112.03769}}.

\bibitem{ParticleDataGroup:2022pth}
{\scshape Particle Data Group} collaboration, \emph{{Review of Particle
  Physics}}, \href{https://doi.org/10.1093/ptep/ptac097}{\emph{PTEP} {\bfseries
  2022} (2022) 083C01}.

\bibitem{Zhang:2016zsp}
Z.~Zhang, \emph{{Time to Go Beyond Triple-Gauge-Boson-Coupling Interpretation
  of $W$ Pair Production}},
  \href{https://doi.org/10.1103/PhysRevLett.118.011803}{\emph{Phys. Rev. Lett.}
  {\bfseries 118} (2017) 011803}
  [\href{https://arxiv.org/abs/1610.01618}{{\ttfamily 1610.01618}}].

\bibitem{Kamenik:2023ytu}
J.~F. Kamenik, A.~Korajac, M.~Szewc, M.~Tammaro and J.~Zupan, \emph{{Flavor
  violating Higgs and $Z$ decays at FCC-ee}},
  \href{https://arxiv.org/abs/2306.17520}{{\ttfamily 2306.17520}}.

\bibitem{Fraser:2018ieu}
K.~Fraser and M.~D. Schwartz, \emph{{Jet Charge and Machine Learning}},
  \href{https://doi.org/10.1007/JHEP10(2018)093}{\emph{JHEP} {\bfseries 10}
  (2018) 093} [\href{https://arxiv.org/abs/1803.08066}{{\ttfamily
  1803.08066}}].

\bibitem{Cranmer:2015bka}
K.~Cranmer, J.~Pavez and G.~Louppe, \emph{{Approximating Likelihood Ratios with
  Calibrated Discriminative Classifiers}},
  \href{https://arxiv.org/abs/1506.02169}{{\ttfamily 1506.02169}}.

\bibitem{Alwall:2011uj}
J.~Alwall, M.~Herquet, F.~Maltoni, O.~Mattelaer and T.~Stelzer, \emph{{MadGraph
  5 : Going Beyond}},
  \href{https://doi.org/10.1007/JHEP06(2011)128}{\emph{JHEP} {\bfseries 06}
  (2011) 128} [\href{https://arxiv.org/abs/1106.0522}{{\ttfamily 1106.0522}}].

\bibitem{Frixione:2021zdp}
S.~Frixione, O.~Mattelaer, M.~Zaro and X.~Zhao, \emph{{Lepton collisions in
  MadGraph5\_aMC@NLO}},  \href{https://arxiv.org/abs/2108.10261}{{\ttfamily
  2108.10261}}.

\bibitem{Sjostrand:2007gs}
T.~Sjostrand, S.~Mrenna and P.~Z. Skands, \emph{{A Brief Introduction to PYTHIA
  8.1}}, \href{https://doi.org/10.1016/j.cpc.2008.01.036}{\emph{Comput. Phys.
  Commun.} {\bfseries 178} (2008) 852}
  [\href{https://arxiv.org/abs/0710.3820}{{\ttfamily 0710.3820}}].

\bibitem{deFavereau:2013fsa}
{\scshape DELPHES 3} collaboration, \emph{{DELPHES 3, A modular framework for
  fast simulation of a generic collider experiment}},
  \href{https://doi.org/10.1007/JHEP02(2014)057}{\emph{JHEP} {\bfseries 02}
  (2014) 057} [\href{https://arxiv.org/abs/1307.6346}{{\ttfamily 1307.6346}}].

\bibitem{Chen:2017yel}
C.~Chen, X.~Mo, M.~Selvaggi, Q.~Li, G.~Li, M.~Ruan et~al., \emph{{Fast
  simulation of the CEPC detector with Delphes}},
  \href{https://arxiv.org/abs/1712.09517}{{\ttfamily 1712.09517}}.

\bibitem{Cacciari:2008gp}
M.~Cacciari, G.~P. Salam and G.~Soyez, \emph{{The anti-$k_t$ jet clustering
  algorithm}}, \href{https://doi.org/10.1088/1126-6708/2008/04/063}{\emph{JHEP}
  {\bfseries 04} (2008) 063} [\href{https://arxiv.org/abs/0802.1189}{{\ttfamily
  0802.1189}}].

\bibitem{Ellis:2019zex}
J.~Ellis, S.-F. Ge, H.-J. He and R.-Q. Xiao, \emph{{Probing the scale of new
  physics in the $ZZ\gamma$ coupling at $e^+e^-$ colliders}},
  \href{https://doi.org/10.1088/1674-1137/44/6/063106}{\emph{Chin. Phys. C}
  {\bfseries 44} (2020) 063106}
  [\href{https://arxiv.org/abs/1902.06631}{{\ttfamily 1902.06631}}].

\bibitem{Ellis:2020ljj}
J.~Ellis, H.-J. He and R.-Q. Xiao, \emph{{Probing new physics in dimension-8
  neutral gauge couplings at e$^{+}$e$^{?}$ colliders}},
  \href{https://doi.org/10.1007/s11433-020-1617-3}{\emph{Sci. China Phys. Mech.
  Astron.} {\bfseries 64} (2021) 221062}
  [\href{https://arxiv.org/abs/2008.04298}{{\ttfamily 2008.04298}}].

\bibitem{kingma2017adam}
D.~P. Kingma and J.~Ba, \emph{Adam: A method for stochastic optimization},
  2017.

\bibitem{caruana2000overfitting}
R.~Caruana, S.~Lawrence and C.~Giles, \emph{Overfitting in neural nets:
  Backpropagation, conjugate gradient, and early stopping}, {\emph{Advances in
  neural information processing systems} {\bfseries 13} (2000) }.

\bibitem{dietterich2000ensemble}
T.~G. Dietterich, \emph{Ensemble methods in machine learning},  in
  \emph{International workshop on multiple classifier systems}, pp.~1--15,
  Springer, 2000.

\end{thebibliography}\endgroup

\end{document}